# Direct Observation of Thermalization to a Rayleigh-Jeans Distribution in Multimode Optical Fibers

Hamed Pourbeyram[1*], Pavel Sidorenko[1*], Fan O. Wu[2], Nicholas Bender[1], Logan Wright[3], Demetrios N. Christodoulides[2], and Frank Wise[1]

[1] School of Applied and Engineering Physics, Cornell University, Ithaca, NY 14853, USA
[2] CREOL/College of Optics and Photonics, University of Central Florida, Orlando, Florida 32816, USA
[3] Physics & Informatics Laboratories, NTT Research, Inc., 940 Stewart Drive, Sunnyvale, CA 94085, USA

frank.wise@cornell.edu

*These authors contributed equally to this work.

## Abstract

Recent years have witnessed a resurgence of interest in nonlinear multimode optical systems where a host of intriguing effects have been observed that are impossible in single-mode settings[1-9]. While nonlinearity can provide a rich environment where the chaotic power exchange among thousands of modes can lead to novel behaviors, at the same time, it poses a major challenge in terms of understanding and harnessing these processes to advantage. Over the years, statistical models have been developed to macroscopically describe the response of these complex systems[10-17]. One of the cornerstones of these theoretical formalisms is the prediction of a photon-photon mediated thermalization process that leads to a Rayleigh-Jeans distribution of mode occupations[10,11,16,17]. Here we report the use of mode-resolved measurement techniques to make the first direct observations of thermalization to a Rayleigh-Jeans power distribution in a multimode optical fiber. We experimentally demonstrate that the underlying system Hamiltonian remains invariant during propagation while power equipartition takes place among degenerate groups of modes – all in full accord with theoretical predictions. Our results may pave the way toward a new generation of high-power optical sources whose brightness and modal content can be controlled using principles from thermodynamics and statistical mechanics [18].

Nonlinear interactions in many-mode systems are ubiquitous in nature. In condensed matter physics, for example, anharmonic processes are known to limit the thermal conductivity of solids and are behind high-temperature superconductivity in binary alloy systems[19,20]. The role of nonlinearity in lattices with many degrees of freedom was first explored in the 1950s by Fermi *et al.* in an effort to support the ergodic hypothesis during thermalization[21] – a seminal study that ushered in the development of soliton and ergodic theories, chaos, and integrable systems[22]. Appreciation of the profound effects that nonlinearities can have on fields that propagate in multimode structures such as cavities, waveguides and random media[23,24] drives renewed interest in addressing these issues in optical settings. The resurgence of interest in systems that support a multitude of modes is driven by the promise of high-bandwidth communication networks[25] and high-power light sources[9]. Along these lines, a series of experiments has been carried out in multimode silica fibers whereby a number of surprising effects has been observed. Examples include geometric parametric instabilities[3,4,26], broadband supercontinuum[26,27], and new families of spatiotemporal dispersive waves[1,8]. In addition, a somewhat unexpected effect has been independently reported in several studies, where, at high intensities, power is found to redistribute itself in a manner that favors the lower groups of spatial eigenmodes of the fiber: the so-called beam self-cleaning effect[2,26]. This peculiar process has nothing to do with self-focusing or Raman effects; energy is exchanged among modes by four-wave mixing. In view of these developments, a new set of theoretical challenges arose in terms of interpreting and predicting the complex nonlinear response of such systems, especially when a large number of modes is involved. This complexity can be readily appreciated by keeping in mind that in a nonlinear system that supports $M$ modes, one must account for $M^2$ cross-phase modulation products and $M^4$ four-wave mixing pathways through which the modes can exchange energy[28]. The analysis of this class of problems is exceedingly complex, and it is difficult to glean physical insight from computational approaches.

    To address these issues, models have been put forward based on notions from wave turbulence/kinetic theories[10,11] and statistical mechanics[16,17,29]. A central prediction of these theories is that the power in nonlinear multimode systems should irreversibly settle after thermalization (*i.e.,* in thermal equilibrium) into a Rayleigh-Jeans (RJ) distribution, a distribution that is uniquely characterized by an optical temperature $T$ and a chemical potential $\mu$. In general, once the RJ distribution is reached, the power $|c_i|^2$ occupying a mode $i$ is given by $|c_i|^2 = -T/(\beta_i + \mu)$ where $\beta_i$ is the linear propagation constant (eigenvalue) associated with the mode.

We note that the RJ distribution in such multimode arrangements is all-optically induced *via* four-wave mixing (Fig. 1**a**) as the system maximizes its entropy in phase space, while constrained to move on the constant-power ($P = \sum_i |c_i|^2$) and Hamiltonian ($H = \sum_i \beta_i |c_i|^2$) isosurfaces. In high-quality fibers power is preserved during propagation, while the linear Hamiltonian is also a constant of the motion since it physically represents the longitudinal momentum flow of the electrodynamic field[16], as expected from the $T_{zz}$ element of the Maxwell stress tensor[30]. In all cases, the optical temperature and chemical potential represent thermodynamic forces that govern the flow of the longitudinal momentum and power, respectively, in a closed Hamiltonian system[18]. The wave-turbulence description of wave condensation – a process found to be related to beam self-cleaning[2] – requires the introduction of a frequency cutoff to avoid divergence of the energy for condensation: Aschieri *et al.* pointed out that a parabolic-index waveguide provides such a cutoff [31]. In the last two years, experimental efforts have exposed some of the ramifications of the expected RJ distribution in multimode optical fibers[6]. From near-field and far-field intensity profiles measured in beam-cleaning experiments, the average occupancy of the fundamental mode was determined[32] during condensation. The corresponding entropy and heat capacity were consistent with equilibrium thermodynamics. In addition, it was shown that the average mode number is conserved in this process[7]. While measurement of the fundamental mode occupancy agrees with the theoretical prediction on average, the distribution cannot be inferred by monitoring only the fundamental mode.

In this paper, we provide the first unequivocal demonstration of RJ thermalization in a multimode parabolic optical fiber. Modal decomposition of the directly-measured, complex, electric field[33] reveals how the modal groups are populated throughout the thermalization process. For some initial conditions, the thermal distribution can be reached after a few characteristic nonlinear lengths of propagation. We find that power equipartition takes place among degenerate groups of modes, in accord with theoretical expectations. Importantly, we find that the output beam quality is invariant, even after the onset of beam self-cleaning.

The experimental arrangement used to observe the thermalization processes is illustrated in Fig. 1**b**. A fiber laser[9] generates 200-fs pulses at 1040 nm that consist of multiple transverse modes locked together in time. With pulse energy of about 20 nJ, the peak power can be as high as 70 kW. The pulses are launched into various lengths of graded-index (GRIN) multimode fiber. Unless specified otherwise, the fiber has a 50-μm core diameter, numerical aperture NA=0.2, and zero-dispersion wavelength near 1300 nm. GRIN fiber is chosen because its equally-spaced propagation constants facilitate four-wave mixing interactions among modes (Supplementary Information). This fiber supports about 220 spatial eigenmodes (in both polarizations) at 1040 nm. Different combinations of transverse modes can be excited by controlling the position and angle of the lens that couples light into the fiber while maintaining 80% coupling efficiency, and losses are negligible. While the input field is linearly polarized, the polarization changes as the pulse accumulates nonlinear phase. The output is projected onto a basis of orthogonal linear polarizations (Methods). For each polarization, the complex electric field is measured by off-axis digital holography and decomposed into a spectrally-resolved eigenmode basis (see Methods and Supplementary Information). The modal power $|c_i|^2$ is obtained by summing the contributions of the two polarizations. The peak power is used as a control parameter to observe the evolution to a thermal distribution. Each measured modal distribution is an average over ~$10^{10}$ pulses, which exhibit ~1% fluctuations in the peak power. Short lengths of fiber were chosen for the experiments to preclude the possibility of stimulated Raman scattering. With fiber lengths between 0.5 and 2.5 m and bend radii about 0.4 m, linear mode coupling is weak. The differential group delay between modes in the parabolic GRIN fiber used is small compared to the effects of chromatic dispersion. The characteristic (chromatic) dispersion length is $L_{DS}$~0.5 m while the nonlinear length is $L_{NL}$ ~ 5 cm for the highest-energy pulses launched into the fiber (characteristic lengths are defined in the Supplementary Information).

When the launched beam has some overlap with the fundamental mode of the fiber, the process of thermalization can be observed. Typical results obtained in a 50-cm-long segment of fiber, with a 50-μm core diameter, will be described first. The beam profiles and modal decompositions observed at the output of the fiber are shown in Fig. 2 for the indicated range of input peak powers. The total power conveyed in each mode group is averaged over its degeneracy $g_k$ to plot the mode occupancy *versus* propagation constant. Because linear mode coupling is weak, the distribution in the top panel of Fig. 2 reflects the modal composition of the launched field. As the peak power is increased above ~10 kW, the spatial profile becomes an intense bell-shaped lobe with approximately the diameter of the fundamental mode of the fiber, accompanied by a low-intensity background. As the power increases, the measured modal occupancies clearly evolve from the initial nonequilibrium distribution at the input to a thermal-equilibrium state. The modal decomposition is performed for each frequency in the spectrum, where it is appreciable. At a peak power of 52 kW, a RJ distribution is observed. From the initial modal distribution (recorded at an input power of 1 kW), one can directly predict the optical temperature and chemical potential by invoking the universal equation of state $H + \mu P = -MT$ where $M$ represents the number of participating modes and $P = \sum |c_i|^2$ is the total power flowing in the system[16] (Supplementary Information). Here, for convenience, the total power flowing in the system is now normalized to unity. Under these conditions, and from the input excitation data, theory predicts that this system should settle to a RJ distribution with $T = 0.40$ mm$^{-1}$ and $\mu = -72.6$ mm$^{-1}$. Experiments carried out at this power level reveal that indeed the system relaxes to a RJ distribution with $T = 0.42$ mm$^{-1}$ and $\mu = -73$ mm$^{-1}$, in close agreement with theory. We emphasize that the theoretical curve is determined from the initial distribution, and is not a fit to the final distribution. The mode-resolved technique employed in our experiment provides the first irrefutable evidence of the RJ distribution that was theorized in earlier works[10,11,16,17,29].

While the result shown in Fig. 2 is associated with a single input condition, the convergence to a Rayleigh-Jeans distribution is a more-general phenomenon. To demonstrate this, we illuminated a 2.5-m length of parabolic index fiber (62.5-µm core diameter and NA=0.275, supporting 812 modes in both polarizations) with an ensemble of statistically-equivalent random input states (speckled fields) and determined the modal occupancy probability density function (PDF) of the entire ensemble at the output, for different input pulse powers (see Supplementary Information for details). In Fig. 3**a**, we show that for sufficiently high peak power (~20 kW), the ensemble-average modal occupancy PDF converges to a Rayleigh-Jeans distribution, which can be predicted from the ensemble-average PDF of a set of low-power (~2 kW) pulses. This means that for any equivalent random-input condition, the corresponding output modal occupation distribution is statistically converging to a Rayleigh-Jeans distribution for appropriate input powers. An example low-power output pattern is shown in Fig. 3**b** for reference. From the ensemble of low-power pulses we find $\langle H \rangle = \sum_i \beta_i \langle |c_i|^2 \rangle \cong 128$ mm$^{-1}$, which allows us to predict the equilibrium temperature and chemical potential ($T = 0.30$ mm$^{-1}$ and $\mu = -178$ mm$^{-1}$) of the ensuing RJ distribution (red line) that corresponds to the high-power ensemble. Figs. 3**c-f** display measurements of different 20-kW wavelength-resolved output intensity patterns, each generated by a different speckled input. The spatial concentration of the intensity at the center of the fiber core in Figs. 3**c-f**, relative to Fig. 3**b**, reflects the speckle-ensemble's convergence to a RJ distribution.

It is worth noting that the experiment of Fig. 2 was performed under conditions corresponding to a microcanonical ensemble: a single input field pattern was used, with both the power and the Hamiltonian of the system invariant during evolution. On the other hand, the experiment of Fig. 3 is effectively performed under canonical-like conditions[18,34], because ensembles of speckle patterns were used. Each speckle pattern is uncorrelated with the other elements in the ensemble, and its precise properties will fluctuate relative to the ensemble average. The speckle patterns fluctuate but yield on average a modal power distribution, while the value of the Hamiltonian fluctuates in a similar manner with a constant expectation value, as expected in a canonical-like setting[18,34]. In this case, the ensemble-average quantities can be used to predict optical-thermodynamic behaviors. As Fig. 3 shows, the experimental results are in good agreement with the theory, which demonstrates the true thermodynamic nature of the process.

Once a modal distribution reaches RJ equilibrium, the distribution values should not change with further evolution. Experiments were performed with longer fibers and/or higher peak powers, to verify that the process is indeed irreversible. In these experiments the beam from the mode-locked laser was used to excite the fiber with 50-μm core, as in Fig. 2. By increasing the input power, the nonlinear length $L_{NL}$ can be adjusted. For peak powers as high as 70 kW launched into 1.5 m of fiber, $L \sim 30 L_{NL}$. With these parameters the spectrum begins to show some asymmetry, from stimulated Raman scattering (Supplementary Information). As in the experiments of Fig. 2, the occupation of the fundamental $LP_{01}$ mode reaches and retains its equilibrium value after propagating a few nonlinear lengths (Fig. 4a). The normalized mean-square error (NMSE) between the measured and expected RJ distributions decreases in the first few nonlinear lengths and remains approximately constant beyond that point (Fig. 4b). Another signature of RJ thermalization in the microcanonical ensemble is the equipartition of power among modes within the same mode group; $|c_i|^2 = -T/(\beta_i + \mu)$, so modes with the same propagation constant $\beta_i$ should be equally-populated at thermal equilibrium. Plots of the occupancies of modes in each group are displayed in the Supplementary Information. We quantify the variation of mode occupancies by calculating the mean coefficient of variation (CV) for the first 6 mode groups, $CV_{mean} = \frac{1}{6}\sum_{k=1}^{6} \frac{\sigma_k}{\overline{|c_k|^2}}$, where $\overline{|c_k|^2}$ is the mean occupancy of mode group $k$ and $\sigma_k$ is the standard deviation in the modal occupancy. In our experiment, the mean CV decreases from 0.32 to 0.15 once thermalization takes place (Fig. 4c). Thus, the powers within the degenerate modes of the same group tend to equalize, in accord with theory.

Numerical simulations based on coupled nonlinear Schrödinger equations[28] were performed to gain further insight into the thermalization process. Details of these simulations are given in the Methods. The measured low-power modal distribution of Fig. 2 was taken as the input field. After an initial exchange of energy between modes, the occupancy of the fundamental ($LP_{01}$) mode increases and becomes dominant after about 15 cm of propagation (Fig. 5**a**). Beyond 20 cm, little energy is exchanged between modes. The numerical results agree with the measured modal distributions as power is varied. In particular, the distribution corresponding to 52 kW agrees well with both our experimental observations and the theoretically-predicted RJ equilibrium state (Fig. 5**b**). Simulations that demonstrate the evolution to the RJ distribution for a range of input powers and fiber lengths are shown in the Supplementary Information.

As previously indicated, in the microcanonical ensemble the invariance of the system's Hamiltonian $H$ during propagation is of paramount importance in establishing the thermodynamic conditions required to observe the RJ distribution[11,16]. Under quasi-linear conditions, *i.e.*, when the intensity is relatively low as in our experiments, one can directly show that the Hamiltonian invariant ($H$) of the system is dominated by its linear kinetic component (Supplementary Information). We verified that the used fiber has negligible losses due to leaky radiation modes and random position-dependent mode-coupling effects. For a given fiber length, as the input peak power is increased, the output spectrum broadens due to self-phase modulation, but remains symmetric and relatively narrowband (the bandwidth is ~3% of the carrier frequency). We measure the Hamiltonian in two different ways: by direct evaluation of $H = \sum \beta_i |c_i|^2$ from mode-resolved measurements of the output beam, and by measuring the ratio of the second moments of the near- and far-field beam profiles, which is proportional to the Hamiltonian in the case of a graded-index fiber (Supplementary Information). Both measures were found to remain constant over the range of input powers employed in the experiments (Supplementary Information). Conservation of the Hamiltonian implies that while the power is reshuffled during thermalization through four-wave mixing, the total photon momentum per watt is preserved. This is consistent with redistribution of power in both directions in mode space, from intermediate to both lower- and higher-order modes.[7] It is important to stress that while the output beam near-field profile in Fig. 2 seems to improve at 52 kW compared to that at 1 kW, the beam quality ($M^2$ factor[35]) is unchanged. This is expected given that $M^2$ is proportional to the Hamiltonian, which is a constant of the motion.

Of course, broad questions remain regarding the observability and universality of thermalization. Fundamentally, the process is universal. However, the equilibrium distribution may not be reached in every experimental setting before dissipation or other competing processes, such as stimulated Raman scattering, occur. The equidistant eigenvalues in GRIN fiber facilitate phase-matching of the four-wave-mixing processes that exchange power among modes (Supplementary Information) and underlie thermalization, and the presence of even weak disorder is known to accelerate the process[6]. Processes that weaken nonlinear interactions among modes will naturally hinder thermalization. It is more difficult to phase-match four-wave mixing in step-index fiber than in GRIN fiber. Thus, higher power will be required to observe thermalization in step-index fiber (Supplementary Information), and stimulated Raman scattering will be harder to avoid. Multimode hollow core fibers filled with atomic gases could be useful in thermalization experiments since the Raman process will no longer be a limitation. Fibers are only a small subset of multimode optical systems; optical thermodynamics will also apply to cavities, multipass cells, and integrated-optical platforms that support multiple modes.

The direct observation of irreversible thermalization to a RJ distribution in GRIN fiber is an important step in harnessing thermodynamic concepts in multimode optical dynamics. More importantly, it provides a verification of optical thermodynamic theories that are based on entropic principles. While the observation of RJ thermalization was carried out in a fiber setting, we expect that these processes will be observed in a variety of environments involving other degrees of freedom such as frequency, optical angular momentum and spin. Finally, it will be interesting to explore some of the other ramifications of optical thermodynamics. These include the observation of equilibria at negative temperatures, isentropic processes, Joule expansions and optical cooling. In addition to unveiling new physics associated with nonlinear complex systems, the thermodynamic approaches could ultimately inspire the design of a new generation of high-performance light sources.

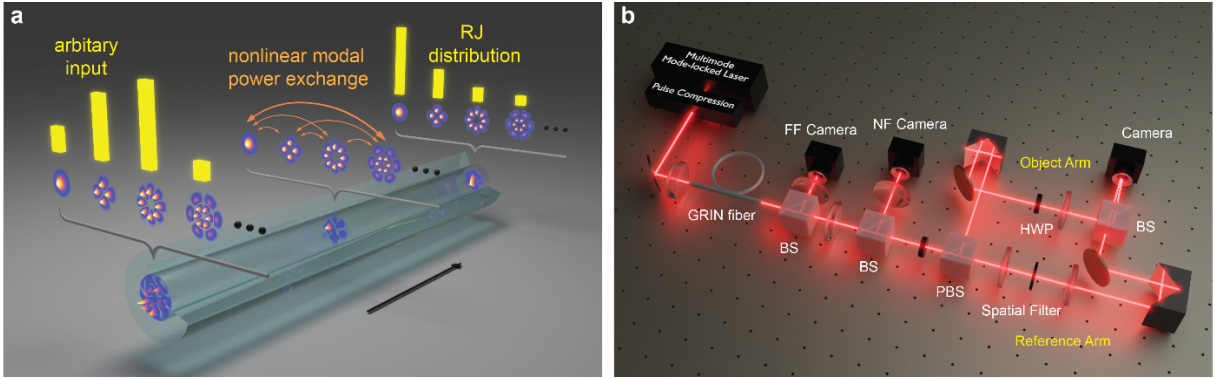

**Fig. 1 | Process of optical thermalization and experimental setup used to observe the Rayleigh-Jeans distribution. a**, A schematic depicting the evolution to optical thermal equilibrium in a nonlinear multimode fiber, arising from four-wave mixing interactions among a multitude of optical guided modes. A superposition of these modes launched into the multimode fiber eventually settles into a Rayleigh-Jeans distribution, thus signifying the onset of thermal equilibrium – in full accord with the second law of thermodynamics. **b**, The experimental arrangement used to demonstrate optical thermalization to a RJ distribution. A multimode mode-locked laser generates femtosecond-duration pulses that consist of multiple transverse modes locked together. The pulses exit the laser with significant frequency chirp, which is removed in the compressor to launch pulses with peak powers up to 70 kW in the multimode GRIN fiber. For experiments with speckle-pattern inputs, a spatial light modulator is inserted between the laser and multimode fiber. The near- and far-field profiles of the output beam are then recorded, and the beam is sent to the interferometric off-axis digital holography setup for spatiotemporal characterization. BS: beam splitter; NF: near field; FF: far field; PBS: polarizing beam splitter; HWP: half-wave plate.

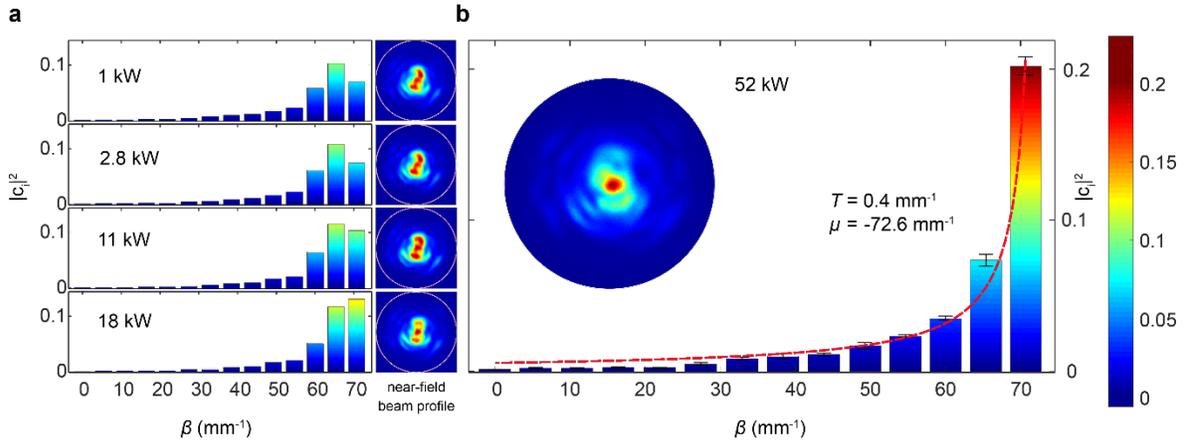

**Fig. 2 | Experimental observation of optical thermalization to a Rayleigh-Jeans distribution. a**, Output power distributions among the GRIN fiber mode groups and their corresponding output near-field intensity profiles, as the total input power increases. Circles in the insets are the core-cladding boundary. The modes are interrogated using interferometric off-axis digital holography. The distributions are plotted against the differential propagation constant, which for the $k^\text{th}$ mode is defined as $\beta_k = \bar{\beta}_k - k_0 n_{cladding}$. The distribution at 1 kW (lowest power level) represents the relative power levels initially launched in the various mode groups at the input. As the power increases, the modal occupancies $|c_i|^2$ are nonlinearly reshuffled. **b**, Direct observation of a RJ distribution upon optical thermalization, which occurs at 52 kW. This thermal equilibrium RJ state, a product of entropy maximization, is characterized by a temperature $T = 0.42 \text{ mm}^{-1}$ and a chemical potential $\mu = -73 \text{ mm}^{-1}$. Error bars represent uncertainties as determined in the Supplementary Information. The dashed curve depicts the theoretically-expected RJ distribution. The discrepancy from the measured distribution for higher-order modes is attributed to the deviation from an ideal parabolic index profile (Supplementary Information). In this same figure, the beam self-cleaning effect in the near field is also apparent in the inset. In all cases, the distributions are plotted in a normalized fashion (*i.e.*, $\sum g_k |c_k|^2 = 1$) where $g_k = k$ represents the mode group degeneracy while the 2-fold polarization degeneracy has been implicitly accounted for. These experiments were conducted in 0.5 m of GRIN fiber with a 50-µm diameter core.

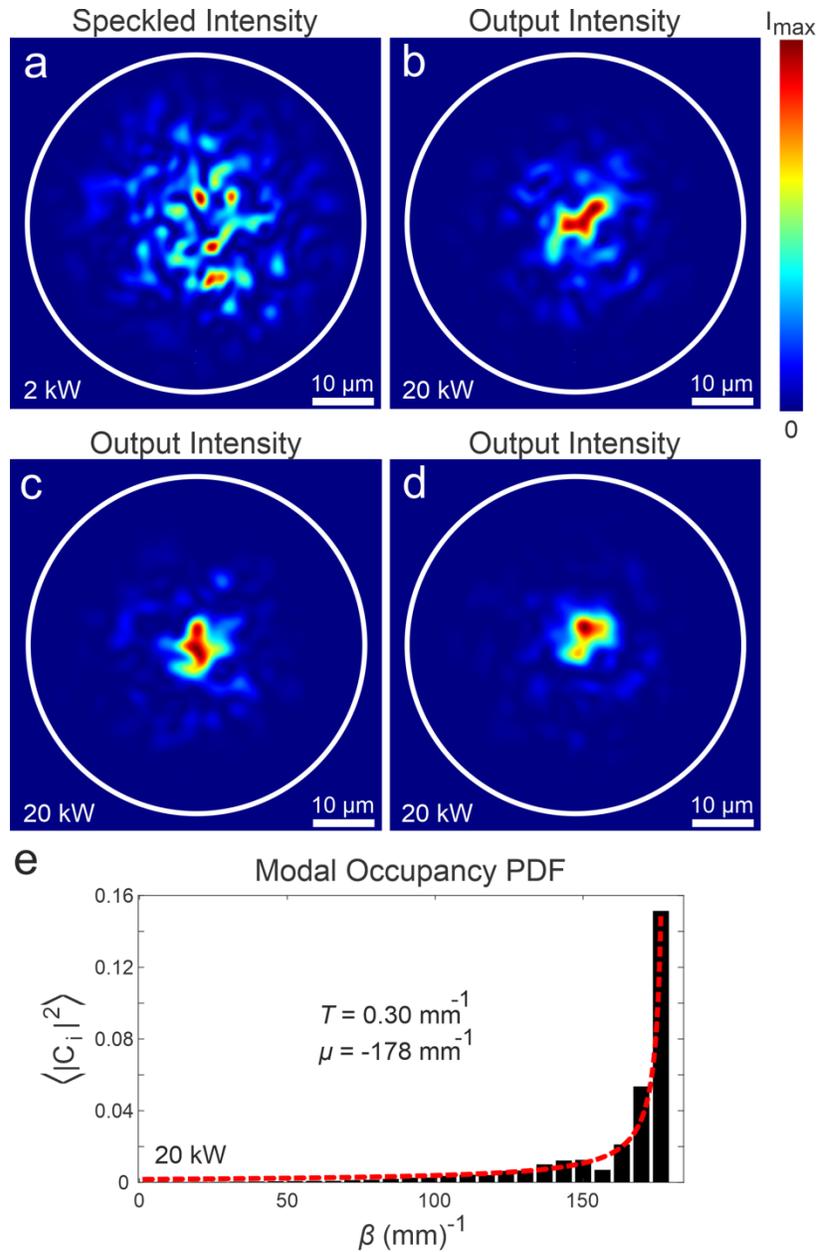

**Fig. 3 | Rayleigh-Jeans ensemble produced by speckled input fields. a)** a typical low-power speckle pattern observed at the output of 2.5 m of GRIN fiber with 62.5-μm core diameter. **b-d)** example intensity profiles observed at the output of the fiber with peak input power of 20 kW. Beam self-cleaning is evident in the profiles. **e)** Observed RJ distribution after averaging over an ensemble of input speckle patterns (similar to the pattern in **a**), at 20 kW. The ensemble-average output modal-occupancy PDF, generated by 14 independent speckled input pulses, is represented by the black bars. The red dashed line is the theoretical RJ distribution obtained from the initial conditions. These results were obtained at a wavelength of 1030 nm.

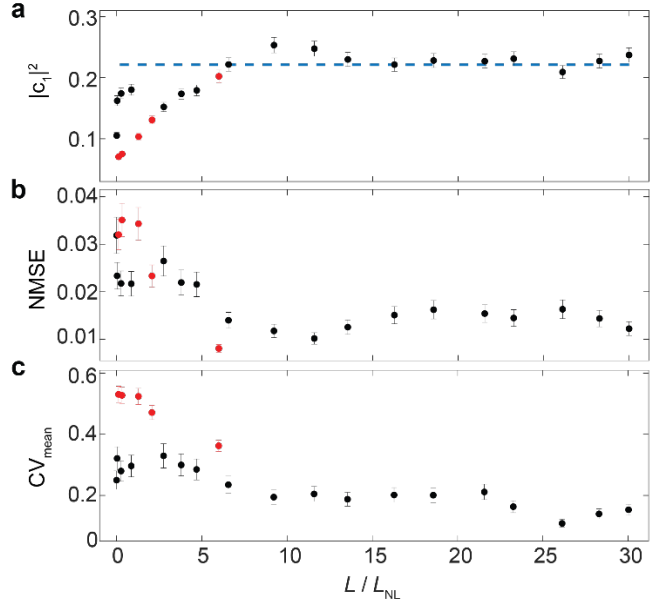

**Fig. 4 | Irreversibility of optical thermalization.** The figures provide information about various signatures of optical thermalization in a nonlinear multimode GRIN fiber, before and after the system attains a RJ distribution. In our experiments the fiber length $L$ is fixed, and the unfolding of thermalization is monitored as a function of $L/L_{NL}$ where the nonlinear length $L_{NL}$ decreases with increasing input power. Red and black symbols represent measurements obtained in a 0.5 m and 1.5 m long fiber, respectively. Error bars represent uncertainties as determined in the Supplementary Information. **a,** Power residing in the fundamental mode $LP_{01}$. The dashed blue line denotes the power occupancy of the fundamental mode, as expected from theory, once the RJ distribution is established. While in the short (0.5 m) fiber, thermal equilibrium is obtained at the highest power level (~52 kW), in the longer (1.5 m) segment the onset of RJ thermalization is established at ~17 kW and henceforth remains at this level. **b,** Normalized mean-square error (NMSE), a measure of the deviation from the theoretical anticipated RJ distribution, as a function of $L/L_{NL}$. As in **a**, the NMSE is irreversibly minimized at the same power levels for the corresponding fiber lengths. **c,** Mean coefficient of variation ($CV_{mean}$) within the first 6 mode groups. Upon thermalization, the $CV_{mean}$ is reduced from 0.32 to 0.15 in the 1.5 m fiber; the standard deviation of the modal occupancies among modes within the same group is suppressed. This signifies equipartition of power within the degenerate modes belonging to the same group, as expected from the RJ distribution. These experiments were conducted in GRIN fiber with a 50-µm diameter core.

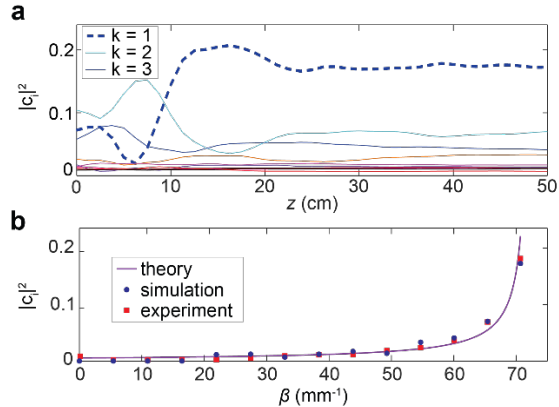

**Fig. 5 | Numerical simulations of the optical thermalization process leading to a Rayleigh-Jeans distribution. a,** Normalized modal power occupancies per mode in each degenerate group $k$ as obtained from numerical simulations carried out in a 0.5 m GRIN fiber with 52 kW input power. The simulations involve the first 10 mode groups and the resulting coupled nonlinear Schrödinger equations were solved using a massively-parallel algorithm. In **a**, the power evolution in the first three mode groups is labeled. **b,** Comparison of the numerical results depicted in **a**, the theoretically predicted thermal RJ distribution, and the experimental results shown in Fig. 2**b**.

## Data availability

All data in the manuscript and supplementary information are available from the corresponding author on reasonable request.

## Code availability

The principal components of the codes used in the manuscript have been made publicly available along with extensive documentation at https://github.com/WiseLabAEP/GMMNLSE-Solver-FINAL.

**Acknowledgements**

This effort was sponsored, in part, by the Department of the Navy, Office of Naval Research under ONR award numbers N00014-20-1-2789, N00014-20-1-2789 and N00014-18-1-2347. Portions of the work were sponsored by the National Science Foundation (ECCS-1912742, EECS-1711230), the Army Research Office (award number W911NF1710481), the Simons Foundation (733682), and the BSF (2016381).

**Author Contributions**

H.P., P.S., and N.B. conducted experiments and performed data analysis. P.S. developed the spatiotemporal measurement instrument. H.P. performed numerical simulations. H.P. and P.S. contributed equally to this work. F.O.W. and D.N.C. developed the theory. L.G.W., H.P., F.O.W. and F. W. conceived of the experiment. All authors contributed to the writing and editing of the manuscript.

**Methods**

**Mode-resolved measurements of the electric field**

Several techniques for measuring the spatial mode distribution of short light pulses have been reported[35-38]. However, measurement of spatiotemporally-complex fields is still quite challenging, especially for fields with large space-time-bandwidth product. For characterization of complex pulsed states, a technique with readily-adjustable space-time-bandwidth product is desirable. Following an approach developed for characterization of intense ultrashort pulses at a focus[39], we employ delay-scanned off-axis digital holography. The pulsed field to be measured is split. One of the copies is filtered spatially to create a Gaussian beam that is used as a spatiotemporal reference beam, and the other (object) is imaged into the camera (Fig. 1**b**). Fringes caused by interference of the reference and object fields contain information about the phase of the latter. The electric field $E(x, y, \tau)$ is Fourier-transformed to obtain $\widetilde{E}(x, y, \omega)$ for the mode decomposition[40-42]. The spatially-filtered reference beam contains all wavelengths in the original field. At each frequency $\omega_i$, the complex field $\widetilde{E}(x, y, \omega_i)$ is projected onto the eigenmode basis of the fiber to get the occupancy of the modes. The relative uncertainty in the mode occupancies is estimated to be 10% (Supplementary Information). As an example, the frequency-integrated modal decomposition of pulses from one particular state of the mode-locked laser is shown in Fig. S5. The beam profile reconstructed from the 3D field modal decomposition is compared to the measured beam profile to assess the mode decomposition.

In general, results for individual polarizations differ. The modal power fractions $|c_i|^2$ plotted in Fig. 2 include both polarizations. The modal power fraction $|c_{ix}|^2$ and power $P_x$ are measured for one polarization. The process is repeated for the other polarization to yield $|c_{iy}|^2$ and power $P_y$. The two polarizations are orthogonal so $|c_i|^2 = (P_x |c_{ix}|^2 + P_y |c_{iy}|^2) / (P_x + P_y)$.

**Numerical simulations**

A system of coupled nonlinear-Schrodinger equations for the temporal envelopes of the spatial modes is solved. The equations include chromatic and modal dispersions, the Kerr and Raman nonlinearities, and self-steepening. The coupled equations are solved with a massively-parallel algorithm on a machine with graphical processing unit functionality[43]. This combination reduces the execution time for relevant simulations of multimode pulse propagation by about two orders of magnitude compared to a standard split-step propagation code run on a central processing unit.

The first 55 transverse spatial modes (first 10 mode groups) of the fiber were included in the calculations, which corresponds to propagation constants larger than $\beta = 20$ mm$^{-1}$. Neglecting higher-order modes was necessary to keep the computation times reasonable; even with the parallel code the execution times exceed several hours for multiple modes with all relevant short-pulse effects included. We expect the neglect of higher-order modes to be a reasonable approximation given the low observed occupancy of those modes, and their uncertainties, in the experiments. In the numerical simulations 200-fs pulses were launched into 50 cm of fiber with the experimental parameters. Simulations performed with a variety of different input distributions but the same values of chemical potential and temperature were found to evolve to the same equilibrium state.

# Supplementary Information

## Direct Observation of Thermalization to a Rayleigh-Jeans Distribution in Multimode Optical Fibers

Hamed Pourbeyram[1], Pavel Sidorenko[1], Fan O. Wu[2], Nicholas Bender[1], Logan Wright[1], Demetrios Christodoulides[2], and Frank Wise[1]

[1]School of Applied and Engineering Physics, Cornell University, Ithaca, NY 14853, USA
[2]CREOL/College of Optics and Photonics, University of Central Florida, Orlando, Florida 32816, USA
[3]Physics & Informatics Laboratories, NTT Research, Inc., 940 Stewart Drive, Sunnyvale, CA 94085, USA
*frank.wise@cornell.edu*

The supplementary material in this document is organized as follows.

**Section 1** illustrates the eigenmodes of graded-index fiber.

**Section 2** describes the measurement of the electric field in some detail.

**Section 3** describes calibration of the measurement of the output beam in the near field.

**Section 4** describes calibration of the measurement of the output beam in the far field.

**Section 5** explains the estimation of the uncertainty in the measured modal occupancies.

**Section 6** is a discussion of the number of mode groups involved in the observed thermalization in graded-index fiber.

**Section 7** provides the characteristic propagation lengths in the experiments.

**Section 8** shows the output spectrum obtained with the highest-power input and longest fiber with a 50 μm diameter core.

**Section 9** explains how to determine the optical temperature and chemical potential.

**Section 10** describes measurements of thermalization made with speckle-pattern input fields.

**Section 11** contains explicit plots of the occupancy of all modes, which show the trend toward equipartition of power across modes in each group.

**Section 12** has results of numerical simulations of the thermalization process performed for varying input peak power and fiber length.

**Section 13** explains the normalized propagation equation and averaged energy used in this work.

**Section 14** calculates the normalized averaged energy used in this work and determines the average mode number.

**Section 15** describes the measurement of the averaged energy in the laboratory.

**Section 16** is a calculation of the normalized mode area ratio for the different spatial modes.

**Section 17** relates the internal energy to the average mode number.

**Section 18** is a discussion of the beam quality and its relation to parameters of the experiments.

**Section 19** provides the connection between the beam quality and the averaged energy measured in the laboratory.

**Section 20** covers evaluation of the linear and nonlinear components of the Hamiltonian, and provides measurements of the Hamiltonian.

**Section 21** is a discussion of thermalization in step-index fiber.

The **Appendix** describes the convention on the Hermite polynomials used in this work.

## Section 1: Modes of GRIN fiber and four-wave mixing processes

The eigenmodes that correspond to each of the lowest 5 eigenvalues of a GRIN fiber are shown on the left side of Fig. S1. Modes with the same eigenvalue are referred to as degenerate mode groups. Power transfers among modes by four-wave mixing, which is described by coupled equations of the form

$$\frac{\partial A_p}{\partial z} \propto \sum_{(l,m,n)} f_{plmn} A_l A_m A_n^* e^{i(\beta_l + \beta_m - \beta_n - \beta_p)z}$$

The process is only efficient when it is phase-matched, that is, when $\Delta\beta = \beta_l + \beta_m - \beta_n - \beta_p = 0$, which is possible when the propagation constants are equally-spaced. The right side of Fig. S1 illustrates an example set of modes that can participate in such a process. Photons in modes $l$ and $m$ would be annihilated, and photons in modes $n$ and $p$ created.

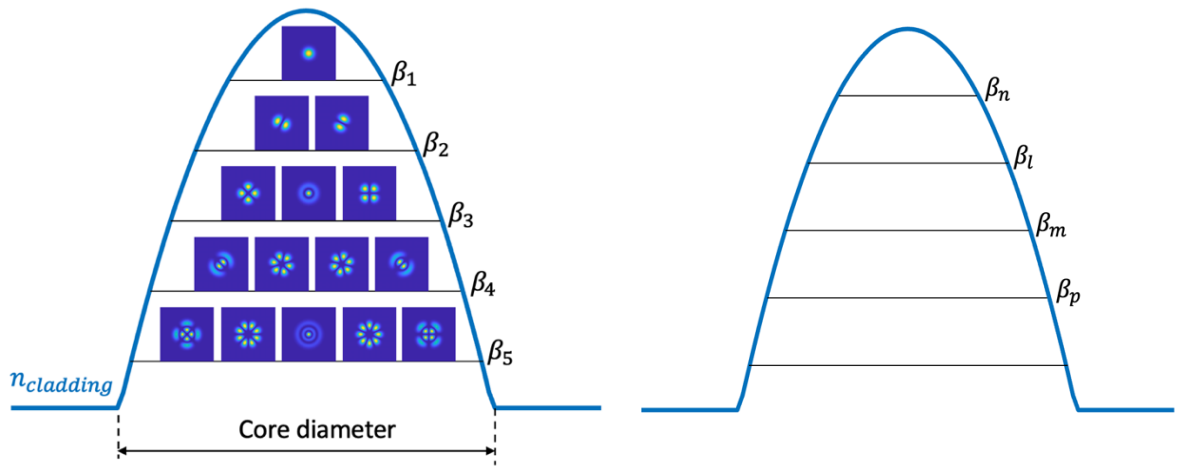

Figure S1. Left: intensity profiles of eigenmodes in the first 5 mode groups of a GRIN fiber. Right: example set of modes that can participate in four-wave-mixing that transfers power among modes.

## Section 2: Mode-resolved measurements

In general, the field that is emitted from a multimode fiber is complex (*i.e.*, it has an amplitude and phase) and depends on position and time, even for a single polarization. For example, it can have different modal content at different times and/or frequencies. In our experiment we perform a complete characterization (amplitude and phase) of the field and then do the mode decomposition for each frequency. The electric field is measured by interferometric off-axis digital holography.

Figure S2 (a) shows the interferometric digital off-axis holography setup, which is an adaptation of the TERMITES method[1]. We analyze two orthogonal polarizations of the incoming spatial-spectral pulse separately. The first pair of HWP-PBS (labeled polarization selector in Fig. S2) at the input of the setup allow us to choose which polarization component will be analyzed. Practically, we perform two independent measurements for two orthogonal linear polarization states of the incoming field. Next, the incoming spatial-spectral pulse is split into two arms: the reference arm and the signal arm. The second pair of HWP-PBS allows adjusting the splitting ratio. The reference arm includes a spatial filter that converts the incoming field to an effective plane wave. The signal arm is delay-scanned and recombined with the reference arm with a non-polarizing beam splitter (NPBS). The half-wave plate before the NPBS is used to ensure the same polarization of the signal and reference arms on the camera. The slight angle between the reference arm and signal arm controls the spacing of the fringes of the interferometric pattern on the camera. Figure S2 (b)-(d) shows the camera frame of only the reference arm, signal arm, and both arms, respectively. Data acquisition is made by scanning the optical path difference (OPD) of the signal arm and acquiring a frame of the camera for each delay step. The scanning is done by motorized piezo stage which is synchronized with the camera trigger. The acquired data can be described by $I(x_i, y_j, \tau_k)$ where $x_i, y_j$ are the ij pixel of the camera and $\tau_k$ is the $k^{th}$ delay.

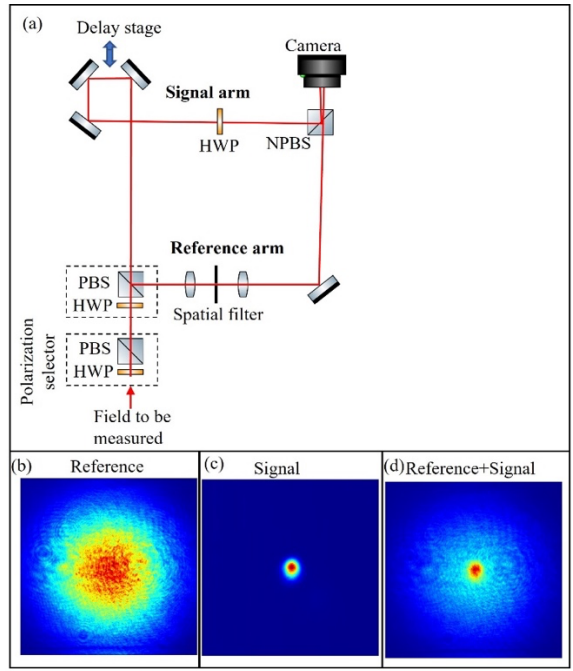

Figure S2. (a) Schematic of the mode resolved measurement setup. PBS - polarizing beam splitter, HWP - half-wave plate, NPBS - nonpolarizing beam splitter. (b) Camera frame of the reference arm only. (c) Camera frame of the signal arm only. (d) Camera frame of the reference and signal arms together.

Next, we convert the acquired data to the frequency domain by performing the Fourier transform on the third dimension $\tau_k$. Figure S3 (a) shows the acquired data at the central pixel of the camera and corresponding Fourier transform. Considering the spectral resolution of our device, the spectrum at the central pixel matches well with the spectrum of the Yb mode-locked laser used in this example. To get the complex field at each frequency, we use the method of off-axis digital holography for each non-zero frequency point from the previous step. Figure S3 (b) illustrates the data analysis for the peak of the spectrum at 291 THz (spectrum presented in Fig. S3 (a) ). The frame that corresponds to the 291 THz is Fourier transformed by 2DFFT. In contrast to conventional holography, only the interference terms are present in the k-space (the DC component was removed when we took only the frame corresponding to the non-zero frequency in the spectrum). We isolate and center one of the peaks in k-space and perform an inverse 2DFFT. The corresponding intensity and phase of the field at frequency 291 THz is shown in Fig. S3 (b).

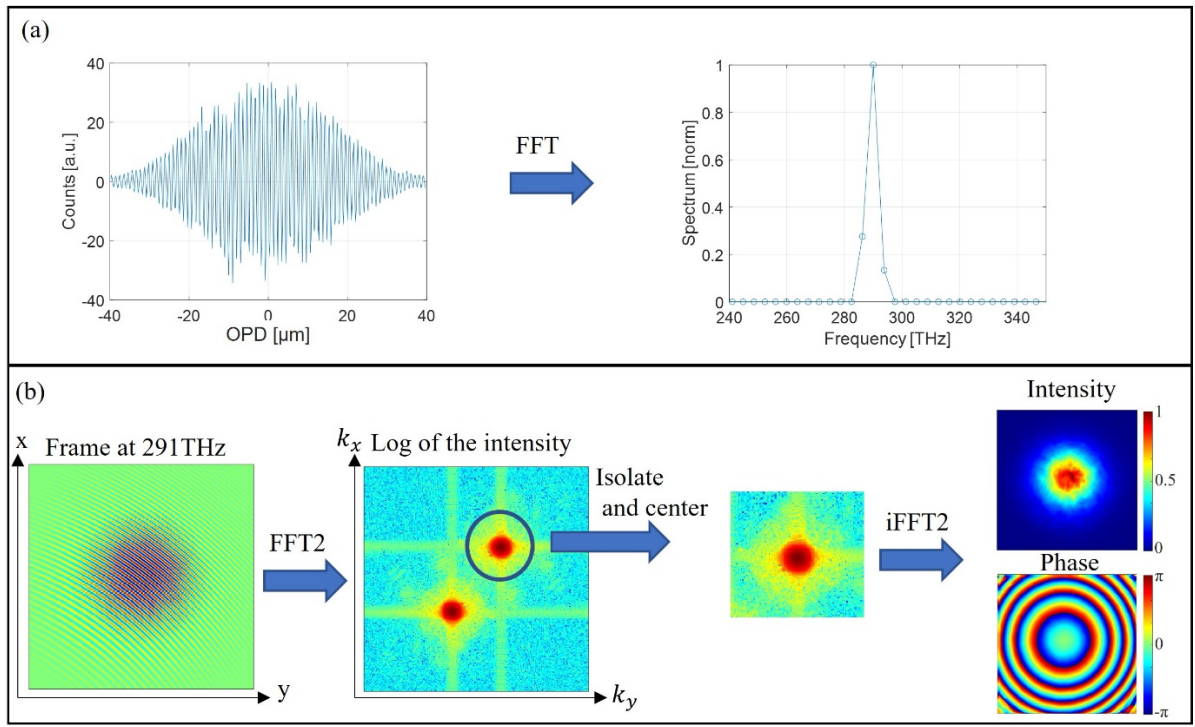

Figure S3. (a) Measured data at the central pixel of the camera and corresponding Fourier transform. (b) Off-axis digital holography data analysis.

While the method described here can be used for amplitude and phase reconstruction of the field at each frequency, the ambiguity of the absolute constant phase term is still present. In other words, with this method, we don't know the phase relation between different frequencies. Nevertheless, this phase relation between different frequencies is not necessary for mode decomposition and it can be determined by performing a FROG measurement on the reference arm of the interferometer[1]. Figure S4 shows a measured case of an inhomogeneous spatial-spectral field. In this example, the reconstructed field is different at different frequencies.

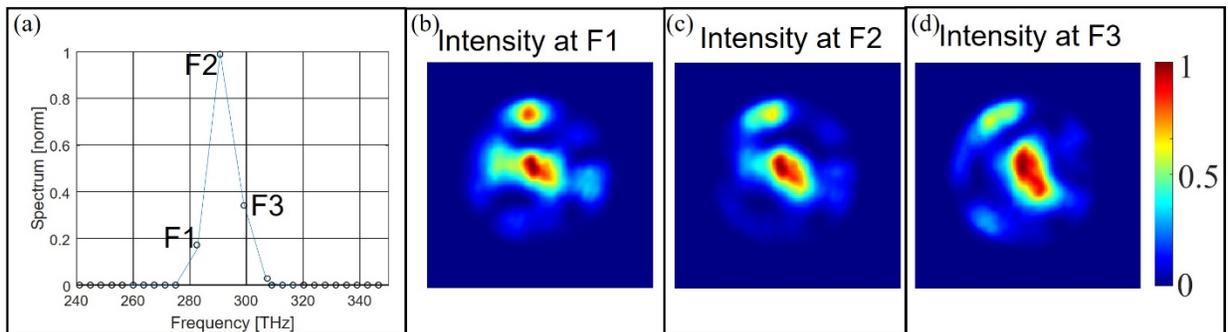

Figure S4. Example of a measured field with spatial-spectral coupling. (a) Spatially integrated power spectrum that was obtained by Fourier transform of the third dimension of the acquired data. (b)-(d) Intensity of the field at corresponding points of the power spectrum.

Finally, the mode decomposition is performed by projection of the reconstructed complex field onto the mode basis. As an example of the input fields used in the experiment, the frequency-integrated modal decomposition of one particular multimode mode-locked state is shown in Fig. S5. 

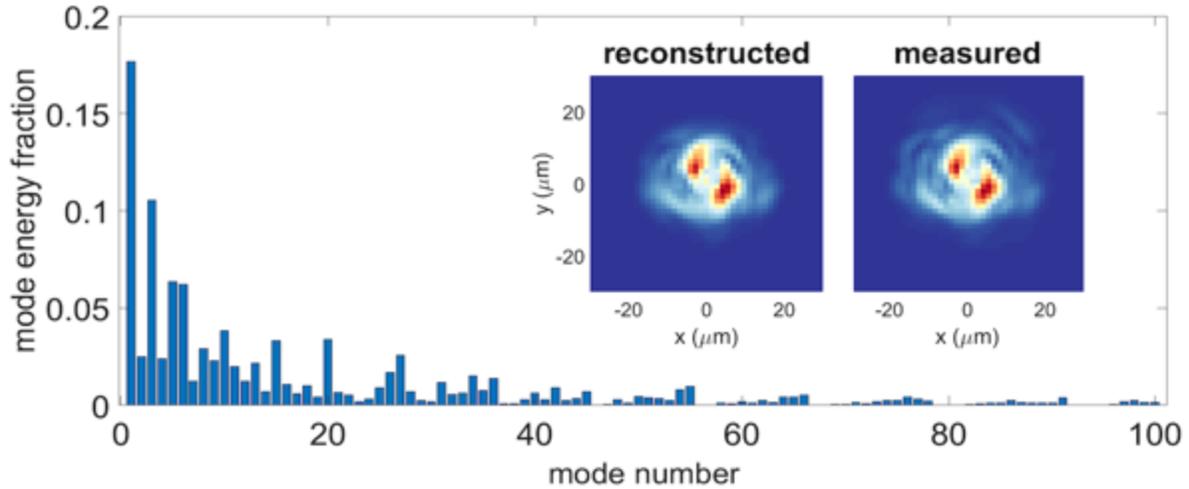

Figure S5. Example of modal decomposition of one mode-locked state of the laser. The intensity profile reconstructed from the modal decomposition is compared to the intensity profile measured directly with a camera.

**Section 3: Calibration of near-field measurement**

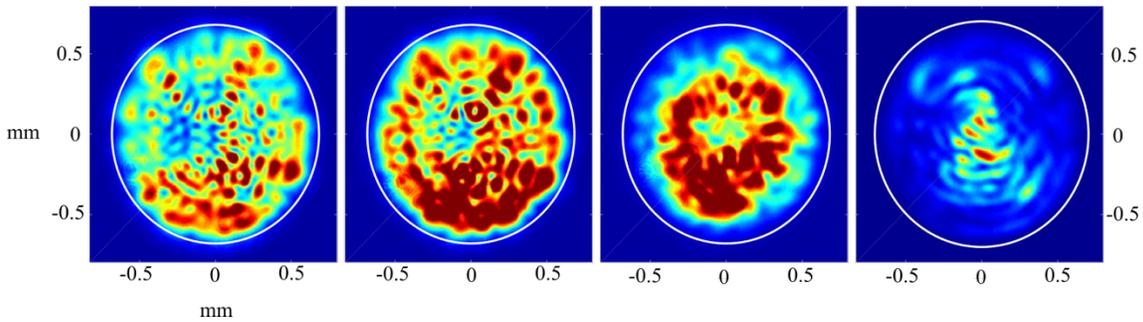

Figure S6. Examples of four different speckle patterns used to calibrate the near-field measurements.

The near-field intensity pattern is measured immediately before each 3D measurement by off-axis digital holography. To calibrate the near-field beam profile measurement, we generate a highly-multimode (MM) speckle pattern by uniformly illuminating the fiber input. In this case, for a highly-MM fiber the near-field pattern approximates the refractive index profile of the fiber core. We use this approach to find the fiber center and approximate radius. We perform this for multiple speckle patterns (Fig. S6) and use the average value for the fiber center. As a check on the calibration we compare the measured core radius with the calculated core radius. The measurements of the circular boundary radius on the CCD camera returns a corresponding fiber core radius of 25.4±0.5 μm, which is in a very good agreement with the nominal value 25±1.25 μm.

**Section 4: Calibration of far-field measurement**

To record the far-field profile we use a thin lens with the same numerical aperture as the GRIN fiber. The fiber and the far-field camera are located in the back and front focal planes of the lens, respectively. To confirm the validity of the measurement and calibrate the measurements we again produce highly-MM speckle patterns (Fig. S7). In this case, the far-field pattern will exhibit a circular shape with radius related to the fiber NA given by $R_{FF} = f_{lens} \times \sin^{-1}(NA)$. Our circular boundary measurements return a corresponding measured NA of $0.175 \pm 0.003$ for the GRIN fiber which agrees well with the known value of $0.200 \pm 0.015$ for the GRIN fiber.

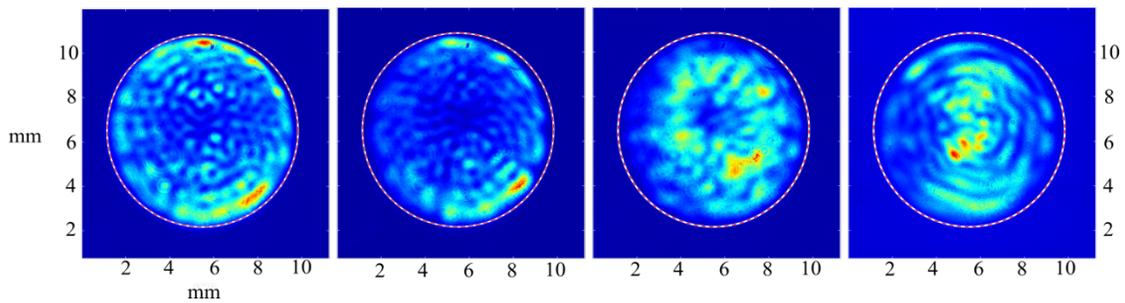

Figure S7. Examples of four different speckle patterns used to calibrate the far-field measurements.

**Section 5: Analysis of uncertainty in mode-resolved measurements**

To assess the uncertainties in the measurements, we compare the modal decomposition obtained with optimum parameters for the fiber center position and transverse scaling with the modal decomposition obtained with the maximum observed deviations from those parameters. The reconstructed and measured profiles for these two limits (an example is shown in Fig. S8) agree quite well. Larger deviations from the optimal parameters will reduce the quality of the reconstruction.

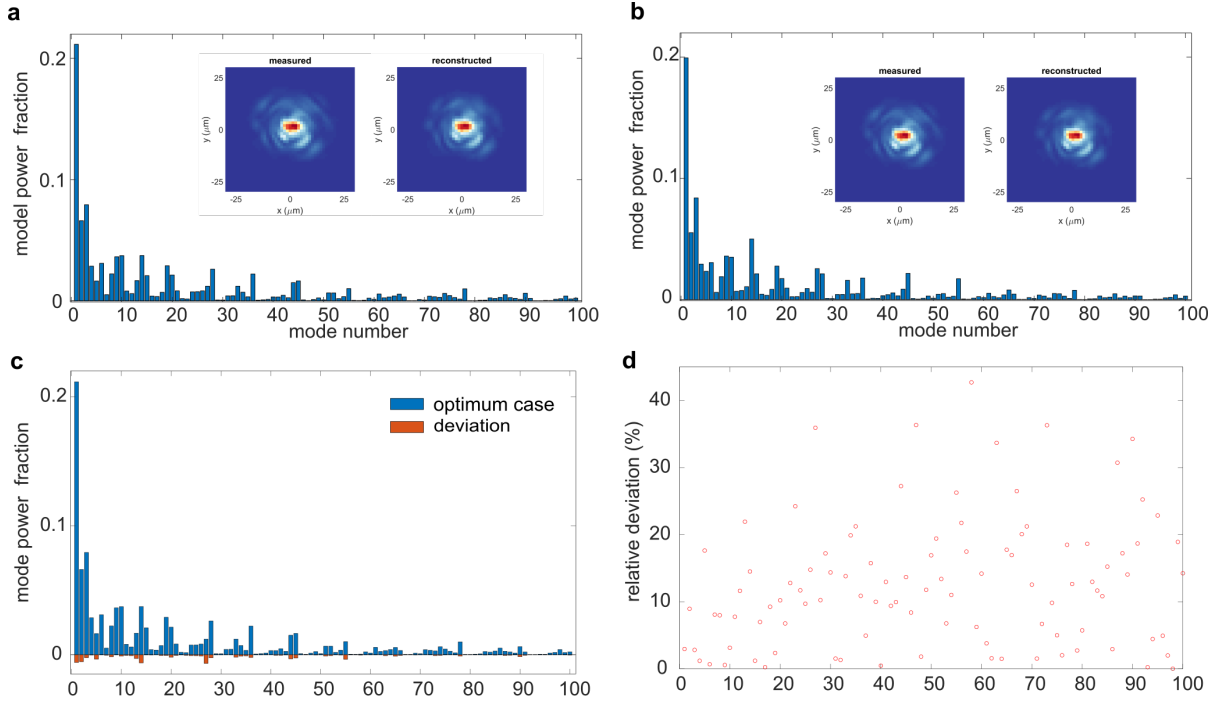

Figure S8. Estimation of uncertainty in modal decomposition. Modal decompositions with the nominal correct image calibration (**a**) and with the fiber center position and transverse scale adjusted by the maximum deviation observed over the course of our experiments, and across multiple methods of obtaining the calibrations (**b**). **c**, blue bars represent the modal occupancies for the nominal correct image calibration (as in panel **a**), and red bars are the magnitudes of the difference between the nominal correct image calibration and that with observed deviation (as in panel **b**). The red bars are plotted with a negative sign for better visual comparison; $-\left||c_i|^2_{opt} - |c_i|^2_{dev}\right|$ is plotted for each mode. **d**, relative uncertainty in the mode energy fraction versus mode.

## Section 6: Mode groups involved in optical thermalization

Optical thermalization is mediated by nonlinear four-wave mixing processes whose efficiency depends on phase matching conditions. In an ideal parabolic potential, the index ladder of the mode groups provides a perfect scenario where the phase matching requirements can be met. However, in practice, the parabolic fiber is truncated by its cladding and hence the index ladder is distorted to some degree, especially for higher-order modes that overlap more strongly with the cladding (Fig. S9). In this respect, nonlinearity plays a crucial role in compensating the four-wave mixing wave vector mismatch via self- and cross-phase modulation, and as a result, it facilitates the nonlinear interactions (energy exchange) between mode groups. On the other hand, thermalization is mediated by chaotic dynamics where the system can access each microstate in a fair manner (i.e., with equal probabilities). In this sense, when the nonlinearity is not strong enough to bridge the deviations from equally-spaced mode indices, modes in higher-order groups no longer participate in the thermalization process. In our 50-µm core experiments, the nonlinear index change caused by a peak power of 50 kW is $\sim 5 \times 10^{-6}$, which is sufficiently strong to bridge the index steps up to mode group 8. Modes belonging to groups 9 and above become isolated because the nonlinearity is not strong enough to compensate the strong departure of eigenvalues from the perfect ladder of a parabolic potential. As a result, in our 50-µm core experiments, only the first 8 groups participate in the thermalization process. Therefore, in estimating the optical temperature and chemical potential at equilibrium, we use $M = 72$, which includes both polarizations in the first 8 mode groups. Note that in our 62.5-µm core fiber, the first 14 mode groups participate in the thermalization process.

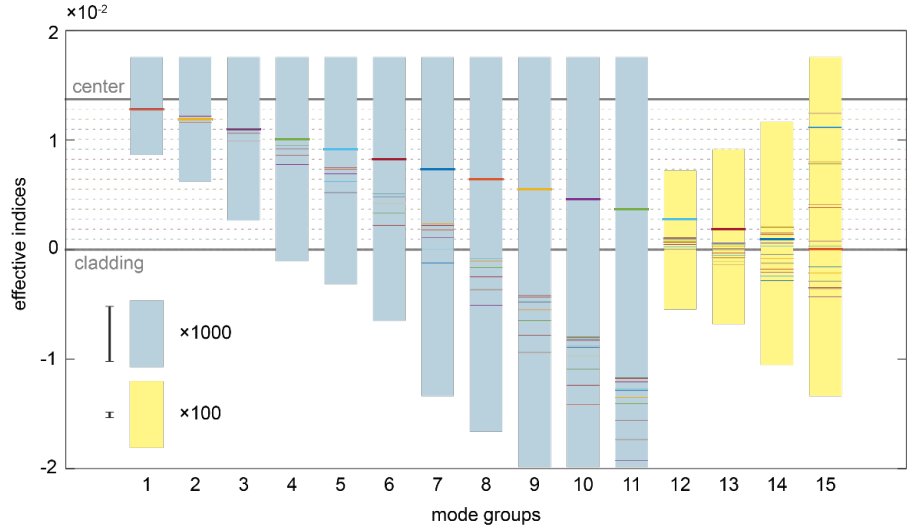

Figure S9. Distribution of effective indices around the values expected for an ideal infinite parabolic potential. The fiber used in our experiment supports totally 240 modes (including both polarizations) that are grouped into 15 levels with a degeneracy $g_k = 2k$ where $k$ represents the mode group number starting from the ground states. The thick solid lines indicate the degenerate mode levels in the corresponding ideal parabolic potential. The thin solid lines represent the effective indices of the actual modes of the fiber used in our experiments, which are obtained using finite element methods. To clearly illustrate these aspects, the eigenvalue distribution has been magnified by a factor of 1000 or 100, as indicated by the legend in the figure. The solid bar represents the nonlinear index change of $5 \times 10^{-6}$ that is required to bridge the mode groups via four-wave mixing. The indices are shifted so the cladding has an index of 0.

**Section 7: Characteristic lengths for nonlinear wave propagation and discussion of spatiotemporal dynamics**

Several characteristic lengths can be defined and compared as a way of estimating the influence of different effects on pulse or beam propagation. The diffraction, dispersion, and nonlinear lengths are given by

$$L_{DF} = \frac{1}{2}k\omega_0^2$$

$$L_{DS} = \frac{T_0^2}{|\beta_2|}$$

$$L_{NL} = \frac{1}{\gamma P_0}$$

where $\omega_0$ is the beam radius, $k$ is the wave vector, $T_0$ is the pulse duration, $\beta_2$ is the group-velocity dispersion, $\gamma$ is an effective nonlinear coefficient and $P_0$ is the peak power of a pulse. In a graded-index fiber, the beam waist of a multimode field oscillates with period $z_p = \pi R/\sqrt{2\Delta}$ where R is the core radius and $\Delta$ is the index difference between the center and the cladding of the fiber. For a beam launched into the core of a multimode fiber, the diffraction length would be a few millimeters. However, the waveguide prevents ordinary diffraction. The multimode field will oscillate or re-image itself with period $z_p$. For the fibers used in our experiments, $z_p \sim 0.5$ mm

Given the use of femtosecond pulses, one might expect spatiotemporal dynamics to play a role in the experiments. However, dispersion does not play a dramatic role in the short fibers used in the experiments described here, and the combination of normal dispersion and modest nonlinear effects is known to lead to gradual temporal broadening of a pulse, as seen in the numerical simulations. The simulations indicate that the beam self-cleaning can occur in a small fraction of the total fiber length (Fig. 5). It is worth mentioning that in contrast to the situation in our experiments, beam-cleaning with nanosecond or sub-nanosecond pulses occurs with complicated spatiotemporal dynamics[2].

**Section 8: Output spectrum with large nonlinear phase accumulation**

The output spectrum that corresponds to the data points at L/L$_{NL}$ = 30 in Fig. 4 is shown in Fig. S10. Note that the y-axis is logarithmic.

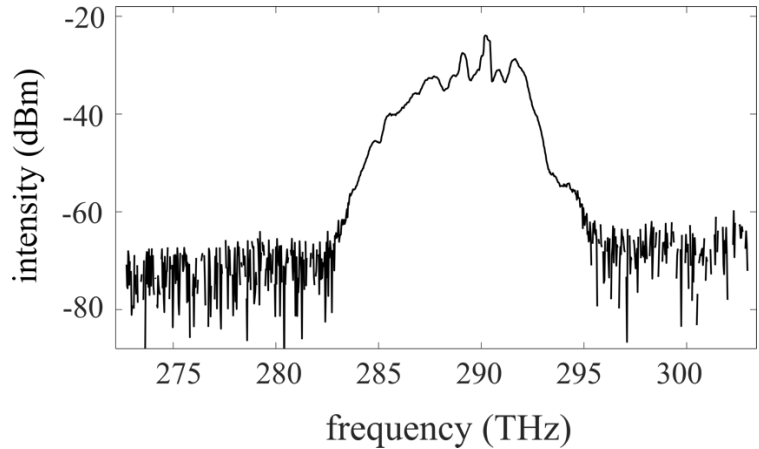

Figure S10. Output spectrum recorded with highest input power and longest fiber with 50-μm core.

**Section 9: Theoretical Prediction of $T$ and $\mu$**

To predict the optical temperature $T$ and chemical potential $\mu$ at thermal equilibrium, the equation of state $H + \mu \mathcal{P} = -MT$ is used to eliminate $\mu$ in the power conservation law. Note that both invariants, i.e., the Hamiltonian $H$ as well as the total power $\mathcal{P}$ can be directly obtained from the initial excitation conditions ($H = \sum_i \beta_i |c_{i0}|^2$ and $\mathcal{P} = \sum_i |c_{i0}|^2$). Once the equation of state is used in the power conservation law, one finds that

$$\mathcal{P} = \sum_i |c_i|^2 = \sum_i \frac{-T}{\beta_i + \mu} = \sum_i \frac{-T}{\left[\beta_i - \frac{(MT + H)}{\mathcal{P}}\right]}.$$

Given that the linear propagation constants of the waveguide $\beta_i$ and the total number of modes $M$ are known, from this algebraic equation, one can directly obtain the temperature $T$ that the system will relax to at equilibrium. The chemical potential can in turn be determined from the equation of state. The uniqueness and existence of the $T$-root of the above equation has been proven in Ref. 3.

**Section 10: Measurements of thermalization with speckle patterns for input**

In this section, an investigation of thermalization performed with speckle-pattern input beams will be described. A GRIN fiber is illuminated by statistically-equivalent ensembles of fully-developed spatially-speckled pulses.

*Experimental Setup*

The experimental setup consists of the arrangement described in Fig. 1 of the main text, with the addition of a reflective phase-only spatial light modulator (SLM; Meadowlark LCoS-1920-1152) to control the illumination pattern incident on the fiber. The SLM is operated in a Fourier transform configuration; in other words, the Fourier-transform of the phase modulation pattern displayed on the liquid crystal screen illuminates the optical fiber. The SLM screen (1920 pixels × 1152 pixels, or 17.6 mm × 10.7 mm) is divided up into 128-pixel by 128-pixel groupings (macro-pixels), resulting in a 15 × 9 macro-pixel phase array, where each macro-pixel is approximately 1.18 mm × 1.18 mm, and the phase modulation of each macro-pixel can arbitrarily be chosen between $0 \leq \theta < 2\pi$. By using macro-pixels to create a uniformly-distributed random phase array on the SLM, a fully developed speckle pattern illuminates the input facet of the fiber. The SLM is illuminated by the de-chirped pulses, produced by the laser source. The laser illumination region on the SLM is circular, with a diameter of approximately 9 macro-pixels. A relay lens in a 2-$f$ configuration images the SLM phase-modulation pattern onto the pupil plane of a 10x objective (LMH-10X-1064), resulting in a circular speckle pattern with an approximate diameter of 20 μm: approximately 9 speckle grains wide. This input illumination pattern is centered relative to the fiber core. For these experiments GRIN fiber with 62.5-μm core diameter and 2.5-m length (Thorlabs GIF-625) was employed. After the fiber, the output pattern is expanded and relayed into the mode-resolved measurement setup (Section 1). The pixel-scale (distance/pixel) of the images recorded by the camera in the interferometer was determined both through selective excitation of the higher-order modes of the fiber (Section 2) and direct measurements of a resolution target (1951 USAF).

*Fiber Modeling*

We use the following fiber properties to model GIF-625: $\alpha = 1.96$, Δn = 1.96%, and a core diameter of 62.5 μm. We find that GIF-625 supports ~ 406 modes (or ~ 28 mode groups) for the wavelengths present in our input light. The modes of GIF-625 are calculated using the same technique as for the GRIN fiber with a 50-μm core.

*Measurements*

Two independent ensembles of speckled-pulses are presented: one with an average peak-power of 2 kW, and one with an average peak power of 20 kW. In the 2-kW ensemble, the spectrally resolved field patterns created by 10 statistically-independent illumination patterns were measured. In the 20-kW ensemble, the spectrally resolved field patterns created by 14 statistically-independent illumination patterns were measured. The difference in the number of patterns is not statistically significant to the following results.

When performing the field measurements with the Mach–Zehnder interferometer, the relative pathlength (between the signal and reference pulses) was stepped/changed sequentially 2001 times, in increments of 0.2 µm per step, for each speckled-pulse illumination of the fiber. After converting the spatially stepped measurement into a temporally delayed measurement, calculating the field at each time delay, and Fourier transforming the resulting fields as a function of time delay, the spectrally-resolved fields can be obtained. To restrict our analysis to non-noisy wavelengths, we only keep the fields associated with the 5 wavelengths with the highest average spectral intensities and only consider the wavelength range where our illumination source is significant (1025 nm – 1045 nm): see Figs. S11 and S12.

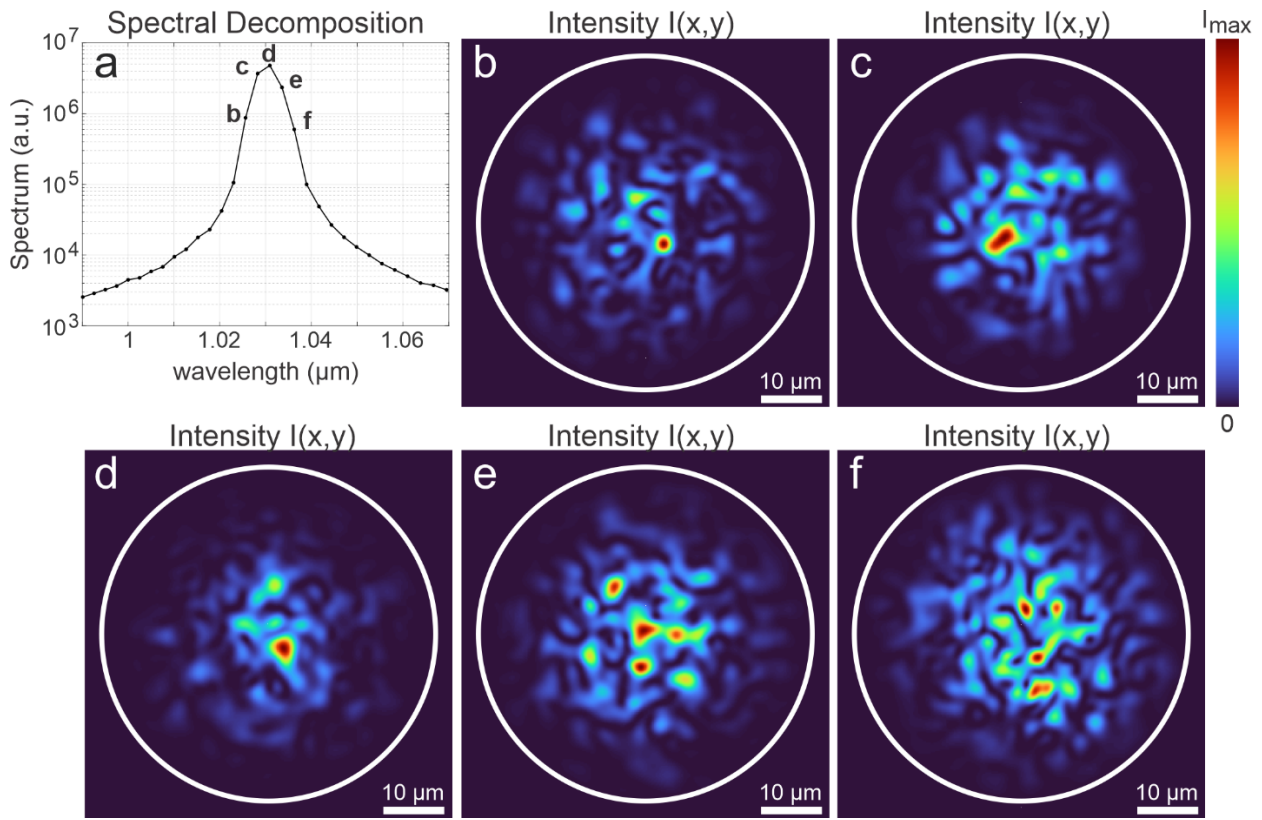

Figure S11: Spectrally resolved fiber outputs, generated by one SLM phase modulation pattern. The measured spectrum **a** along with 5 wavelength resolved speckle patterns **b-f** are shown for one 2 kW pulse. The white circle corresponds to the boundary of the fiber core.

As can be seen in both Fig. S11 and S12, the intensity patterns associated with individual wavelengths are different. Therefore, each speckled pulse generated by the SLM produces 5 patterns which can be modally decomposed. As such, the 2-kW ensemble produces a total of 50 statistical realizations, and the 20 − kW ensemble produces 70 statistical realizations.

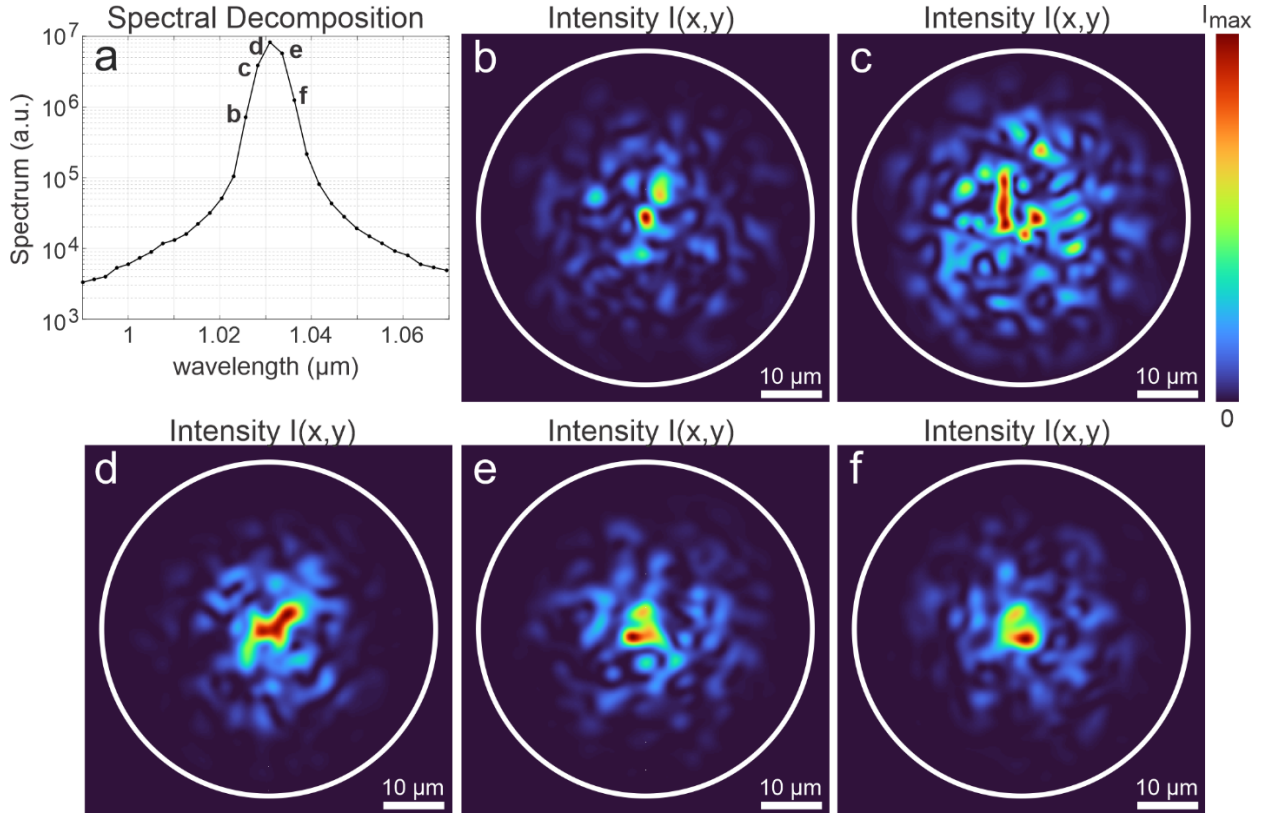

Figure S12: Spectrally resolved fiber outputs, generated by one SLM phase modulation pattern. The measured spectrum **a** along with 5 wavelength resolved speckle patterns **b-f** are shown for one 20 kW pulse. The white circle corresponds to the boundary of the fiber core.

*Data Processing*

The following additional data processing/techniques were used post-measurement and before the mode decomposition.

- For each pulse, the average center of intensity, $\frac{\int I(\bar{r})\bar{r}\,d\bar{r}}{\int I(\bar{r})\,d\bar{r}}$, of the spectrally resolved fields (where the averaging was done over different wavelengths) was used to locate the center of the fiber core in each measurement.
- We correct for aberrations in the phase of the measured field, caused by the slight non-uniformity of the reference pattern.

*Modal Decomposition*

With both the spectrally resolved fields and the numerical predictions for the waveguide modes, a modal decomposition can be performed on the each of the field patterns in a given ensemble. By calculating the complex Pierson correlation coefficients between each of the fiber modes with a measured spectrally-resolved field, we can decompose the measured field into a linear-combination of the waveguide modes, with the norm set to 1. The squared-magnitude of the coefficients in the linear combination gives the normalized modal occupancy of the field-pattern, and averaging over each mode group, $\beta$, gives the modal-group occupancy, $|C_\beta|^2$. These values can be averaged over the different realizations (different spectrally resolved fields and different speckled inputs) in each ensemble to obtain the modal-group occupancy probability density function, $\langle|C_\beta|^2\rangle$, for each ensemble (2 kW, and 20 kW). By calculating the probability density function for different random input patterns, we have access to information about the statistical tendencies of light in the fiber, as opposed to just information about a specific coupling configuration.

*Results*

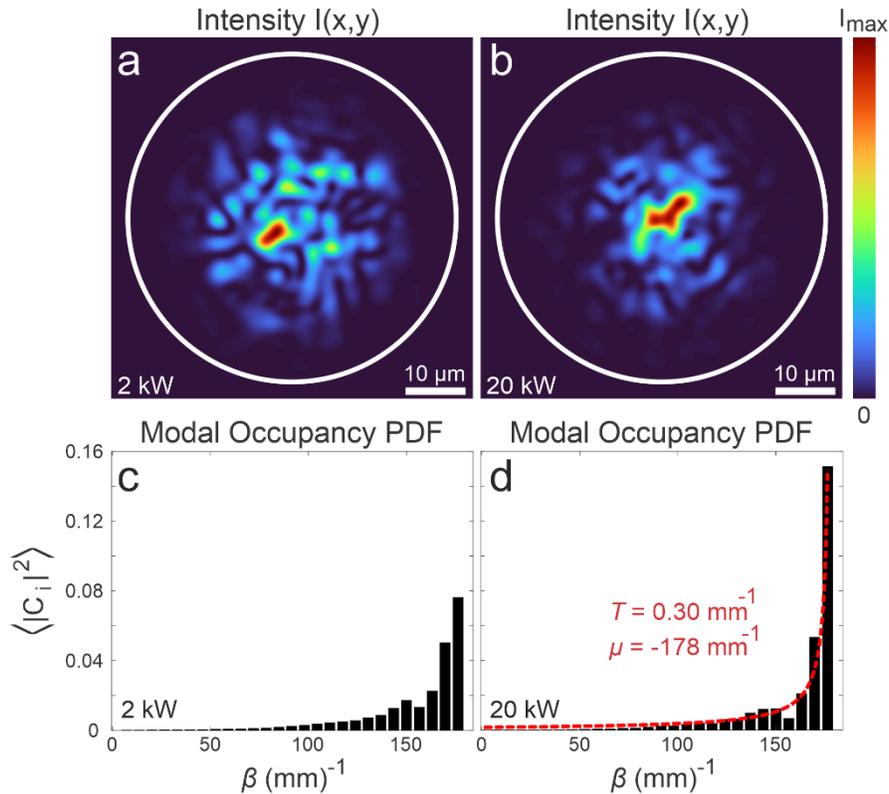

Figure S13: Optical thermalization of an ensemble of speckle patterns. In the top row wavelength-resolved output intensity patterns produced with pulses with an average peak power of 2 kW (**a**) and 20 kW (**b**) are shown. The white circles correspond to the edge of the fiber core. In the second row, the output modal occupancy PDFs generated by ensembles of input speckle patterns corresponding to the examples given in **a,b**, are shown in **c,d**: versus the relative mode propagation coefficient. The temperature and chemical potential of the Rayleigh-Jeans distribution in **d** (red line), are calculated in the same manner as the equilibrium values in Fig. 2 of the main text, using the distribution in **c**.

Two example spectrally-resolved intensity patterns within each ensemble measured (2 kW, and 20 kW) are presented in Figs. S13 **a,b**, along with the corresponding ensemble-average modal occupancy PDFs in Figs. S13 **c,d**, for the entire ensembles. The transition of the PDF, from **c** to **d,** demonstrates the statistical tendency of random-input patterns to transition into the fundamental mode of the fiber a with increasing input power. The speckle pattern in **a** and its PDF **c** represent the low-power behavior of pulsed light through the fiber with 62.5-μm core, given our illumination configuration. While ideally this PDF should be flat, non-uniform mode coupling at the input and modal-dependent loss can prevent this. As such, it is not surprising that for speckle illumination the low-power PDF increases as a function of the propagation constant β. This does *not* imply that all of the speckle patterns in the low-power ensemble are fundamental-mode dominated like the PDF. The different speckle patterns that constitute the ensemble each have mode-group occupancies that fluctuate about the ensemble-average PDF, is some cases relatively starkly (Fig. S14). Nevertheless, the PDF in Fig. S13 **c** can still be used to predict the thermal-equilibrium state in Fig. S13 **d** (red dashed line), $\langle |C_\beta| \rangle = -T/(\beta + \mu)$, where $T = 0.30 \, mm^{-1}$ and $\mu = -178 \, mm^{-1}$. The good agreement between the PDF and red line in Fig. S13 **d** indicates that thermal equilibrium has been reached. Therefore, we have found that for any equivalent random-input condition, the corresponding output modal occupation distribution is statistically-likely to converge to a Rayleigh-Jeans distribution: for appropriate input pulse powers.

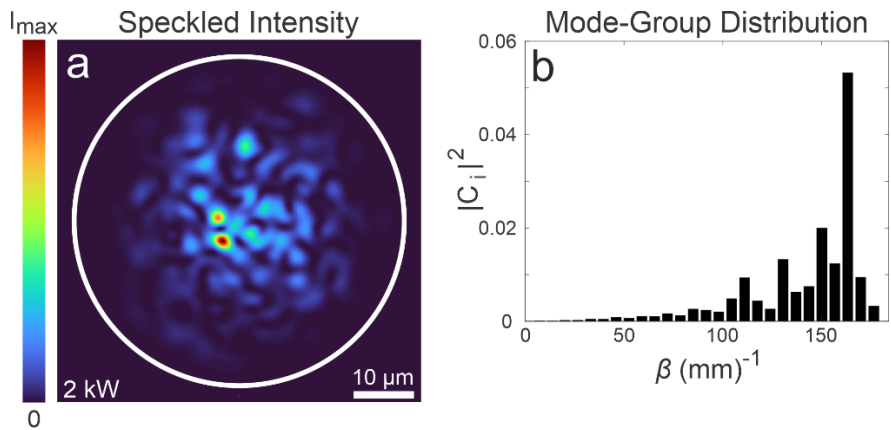

Figure S14: A single low-power speckle pattern **a** along with its mode-group distribution **b**.

## Section 11: Equipartition of power among modes in each mode group

The modal occupancies observed with 50 cm of 50-μm core fiber with 1 kW input power and 52 kW input power are shown in Fig. S15. The modal occupancies observed with 150 cm of 50-μm core fiber with 1 kW input power and 70 kW input power are shown in Fig. S16. In each case, the trend toward equipartition of power across modes in a group is evident.

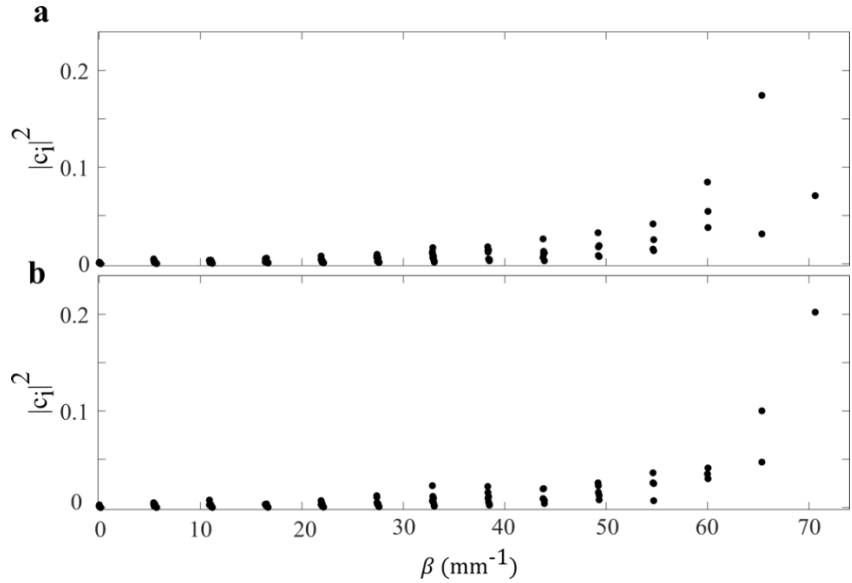

Figure S15. Modal occupancies recorded with 50 cm fiber (50-μm core) at 1 kW (a) and 52 kW (b).

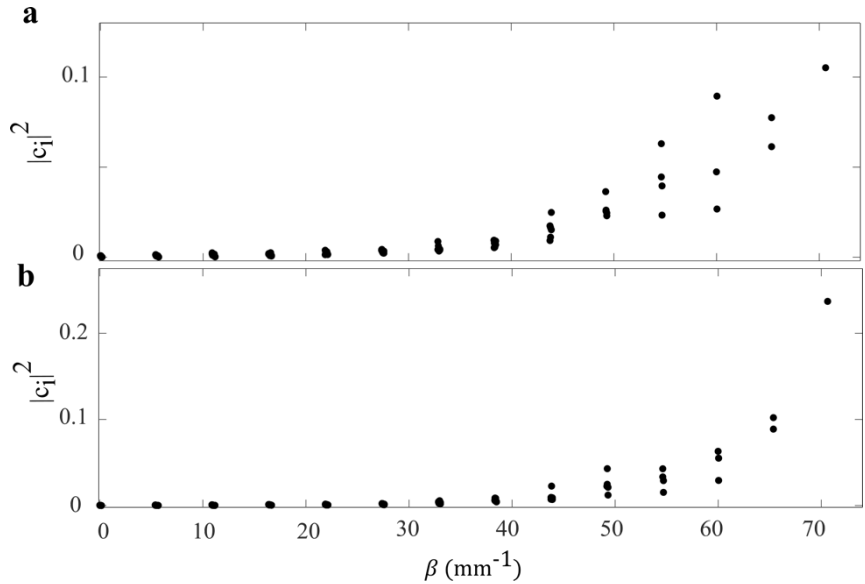

Figure S16. Modal occupancies recorded with 150 cm fiber (50-μm core) at 1 kW (a) and 70 kW (b).

## Section 12: Numerical simulations of thermalization

Simulations show (Fig. S17) that the evolution to the RJ distribution occurs for a range of input peak powers. The input peak power varies from the value at which beam-cleaning and thermalization are first observed clearly, to four times higher. With increasing input power, the process occurs in proportionally-shorter lengths of fiber.

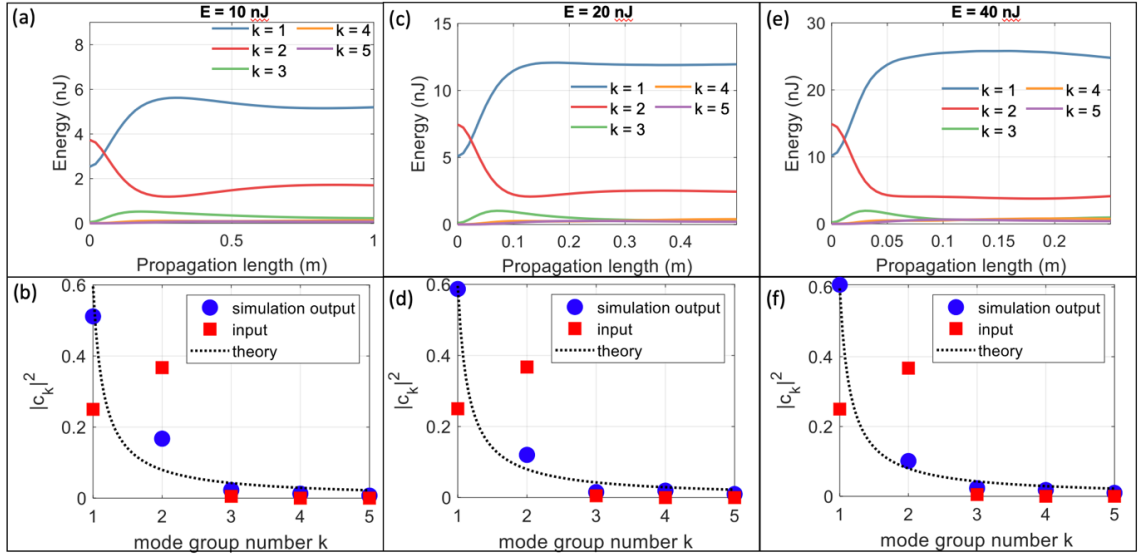

Figure S17. Simulations of thermalization. 200-fs pulses with the indicated pulse energies are launched into a fiber with parameters similar to those used to produce Fig. 5 of the text. Top row: modal occupancies versus propagation distance. Bottom row: initial (red squares), simulated final (blue dots), and theoretical final (dotted line) modal occupancies.

## Section 13: Normalization of propagation equation and average energy $\bar{U}$

In the paraxial regime, the evolution of the optical field $E$ confined in a refractive index potential $n(X, Y)$ is governed by

$$i\frac{\partial E}{\partial Z} + \frac{1}{2k}\frac{\partial^2 E}{\partial X^2} + \frac{1}{2k}\frac{\partial^2 E}{\partial Y^2} + k_0 n(X,Y)E + n_2 k_0 |E|^2 E = 0, \tag{S1}$$

where the propagation constant $k = k_0 n_0$ and $n_2$ is the nonlinear coefficient. By using $Z = 2k_0 n_0 x_0^2 z$, $X = xx_0$, $Y = yx_0$, $E = \rho\psi e^{ik_0 n_0 Z}$ ($n_0$ is the refractive index of the center of the fiber) and $\rho = 1/\sqrt{2k_0^2 n_0 x_0^2 n_2}$, we obtain

$$i\frac{\partial \psi}{\partial z} + \frac{\partial^2 \psi}{\partial x^2} + \frac{\partial^2 \psi}{\partial y^2} + V(x,y)\psi + |\psi|^2 \psi = 0, \tag{S2}$$

where $V(x,y) = 2k_0^2 n_0 x_0^2 [n(x,y) - n_0]$ is the potential energy. In a parabolic potential,

$$[n(x,y) - n_0] = -n_0 \Delta \frac{(x^2 + y^2)x_0^2}{a^2} \tag{S3}$$

where $a$ is the core radius. By setting $x_0 = \left(\frac{a^2}{2k_0^2 n_0^2 \Delta}\right)^{1/4}$, $V$ becomes $V = -(x^2 + y^2)$. Under linear conditions (weak nonlinearity), the above equation can be written as

$$i\frac{\partial \psi}{\partial z} + \nabla_\perp^2 \psi + V(x,y)\psi = 0. \tag{S4}$$

The eigenstates (modes) $\Phi_i = \phi_i e^{i\varepsilon_i z}$ can be found from

$$-\varepsilon_i \phi_i + \nabla_\perp^2 \phi_i + V(x,y)\phi_i = 0. \tag{S5}$$

For an arbitrary optical field $\psi$ satisfying Eq. (S4), multiplying by $\psi^*$ and integrating over transverse coordinates yields

$$-\iint_{-\infty}^{+\infty} i\psi^* \frac{\partial}{\partial z}\psi \, dxdy = \iint_{-\infty}^{+\infty} (\psi^* \nabla_\perp^2 \psi + \psi^* V(x,y)\psi)dxdy = 0. \tag{S6}$$

In general, $\psi$ can be written as a superposition of the $M$ eigenstates in the system,

$$\psi(x,y,z) = \sum_{i=1}^{M} c_i \phi_i(x,y) e^{i\varepsilon_i z}. \tag{S7}$$

Since the eigenmodes are orthogonal, i.e., $\iint_{-\infty}^{+\infty} \phi_i \phi_k^* dxdy = \delta_{ik}$, substituting Eq. (S7) into the left side of Eq. (S6) leads to

$$\sum_{i=1}^{M} \varepsilon_i |c_i|^2 = \iint_{-\infty}^{+\infty} \psi^* \nabla_\perp^2 \psi \, dxdy + \iint_{-\infty}^{+\infty} V(x,y)|\psi|^2 dxdy. \tag{S8}$$

The optical energy $U$ is defined as $U = -\sum_{i=1}^{M} \varepsilon_i |c_i|^2$. Moreover, by converting the first term on the right side of the equation into its Fourier domain, we obtain:

$$U = \iint_{-\infty}^{+\infty} (k_x^2 + k_y^2) |\psi(k_x, k_y)|^2 dk_x dk_y + \iint_{-\infty}^{+\infty} (x^2 + y^2)|\psi(x,y)|^2 dxdy. \tag{S9}$$

The averaged energy is (from this point on, bar on a variable indicates that it is averaged over optical power)

$$\bar{U} = \frac{U}{\mathcal{P}} = \frac{\iint_{-\infty}^{+\infty}(k_x^2 + k_y^2)|\psi(k_x,k_y)|^2 dk_x dk_y}{\iint_{-\infty}^{+\infty} |\psi(k_x,k_y)|^2 dk_x dk_y} + \frac{\iint_{-\infty}^{+\infty}(x^2+y^2)|\psi(x,y)|^2 dxdy}{\iint_{-\infty}^{+\infty}|\psi(x,y)|^2 dxdy}$$

$$= S_{FF} + S_{NF}, \tag{S10}$$

where $S_{FF}$ and $S_{NF}$ are the second moments of the far-field and near-field profiles, respectively.

**Section 14: Averaged energy and averaged mode number**

Consider the fundamental mode

$$\Phi_1 = \frac{\sqrt{v}}{a\sqrt{\pi}} e^{-\frac{(X^2+Y^2)v}{2a^2}} e^{i\varepsilon_1 z} = \frac{\sqrt{v}}{a\sqrt{\pi}} e^{-\frac{(x^2+y^2)vx_0^2}{2a^2}} e^{i\varepsilon_1 z} \tag{S11}$$

where $v = k_0 n_0 a\sqrt{2\Delta}$ is the well-known "V number" of the fiber.
The averaged energy of the fundamental mode is:

$$\bar{U}_1 = \frac{U_1}{\mathcal{P}} = \frac{vx_0^2}{2a^2} + \frac{a^2}{2vx_0^2} = S_{FF}^1 + S_{NF}^1 = 2 \tag{S12}$$

Notice that $x_0^2 v = a^2$. The contribution from the two terms are the same, i.e., $S_{NF}^1 = S_{FF}^1$, and this is true regardless of the normalization applied. The above relation is the general conclusion reached in any harmonic oscillator: the kinetic energy and potential energy are equal. If we further normalize the averaged energy by the fundamental mode, we obtain the so-called averaged mode number $\bar{k}$ (only valid for parabolic fiber):

$$\bar{k} = \frac{\bar{U}}{\bar{U}_1} = \frac{S_{NF} + S_{FF}}{S_{NF}^1 + S_{FF}^1} = \frac{1}{2}\left(\frac{S_{NF}}{S_{NF}^1} + \frac{S_{FF}}{S_{FF}^1}\right) \tag{S13}$$

A similar result has also been provided in Ref. 4.

**Section 15: Measurement of $\bar{U}$**

We can obtain $\bar{U}$ in physical parameters without normalizations from Eq. (S10) by applying $X = xx_0, K_x = k_x x_0, \varepsilon_i z = \beta_i Z$:

$$\bar{U} = -2k_0 n_0 x_0^2 \frac{\sum_{i=1}^M \beta_i |c_i|^2}{\sum_{i=1}^M |c_i|^2} = \frac{a^2}{k_0 n_0 \Delta} \frac{\sum_{i=1}^M \beta_i |c_i|^2}{\sum_{i=1}^M |c_i|^2}$$

$$= x_0^2 \frac{\iint_{-\infty}^{+\infty}(K_x^2 + K_y^2)|\psi(K_x, K_y)|^2 dK_x dK_y}{\iint_{-\infty}^{+\infty}|\psi(K_x, K_y)|^2 dK_x dK_y} + \frac{1}{x_0^2}\frac{\iint_{-\infty}^{+\infty}(X^2 + Y^2)|\psi(X, Y)|^2 dXdY}{\iint_{-\infty}^{+\infty}|\psi(X, Y)|^2 dXdY}$$

$$= \frac{a}{k_0 n_0 \sqrt{2\Delta}} \frac{\iint_{-\infty}^{+\infty}(K_x^2 + K_y^2)|\psi(K_x, K_y)|^2 dK_x dK_y}{\iint_{-\infty}^{+\infty}|\psi(K_x, K_y)|^2 dK_x dK_y} + \frac{k_0 n_0 \sqrt{2\Delta}}{a}\frac{\iint_{-\infty}^{+\infty}(X^2 + Y^2)|\psi(X, Y)|^2 dXdY}{\iint_{-\infty}^{+\infty}|\psi(X, Y)|^2 dXdY} \tag{S14}$$

The two prefactors in Eq. (S14) play important roles similar to that of "natural oscillator units," which has been applied to coherent states in quantum optics.

The average mode number is

$$\bar{k} = \frac{\bar{U}}{\bar{U}_0} = \frac{1}{2}\left(\frac{\iint_{-\infty}^{+\infty}(X^2 + Y^2)|\bar{\psi}(X, Y)|^2 dXdY}{\iint_{-\infty}^{+\infty}(X^2 + Y^2)|\overline{\Phi_1}(X, Y)|^2 dXdY} + \frac{\iint_{-\infty}^{+\infty}(K_x^2 + K_y^2)|\bar{\psi}(K_x, K_y)|^2 dK_x dK_y}{\iint_{-\infty}^{+\infty}(K_x^2 + K_y^2)|\overline{\Phi_1}(K_x, K_y)|^2 dK_x dK_y}\right) \tag{S15}$$

In the lab, we can measure the near-field and far-field intensity patterns, and then the associated second moments are calculated based on the imaging data where the coordinates are unitless. Fortunately, for an arbitrary unit or unitless coordinates used by the camera, as long as the experimental setup is unchanged when measuring the fundamental mode and the thermalized field, those prefactors and units cancel out.

**Section 16: Calculation of $\bar{U}$ for individual spatial modes**

Fig. S18 shows the normalized mode area ratio calculated using the far-field and near-field intensity profiles, for the first 100 modes of a GRIN fiber with 50 μm core, using the commercially available parameters. We calculate the second moments of the near- and far-field patterns, and then divide each value by the second-moment area of the fundamental mode. Each color represents a mode group. The normalized mode area ratio is only a function of the mode group number (g) and in fact is identically equal to the value of g.

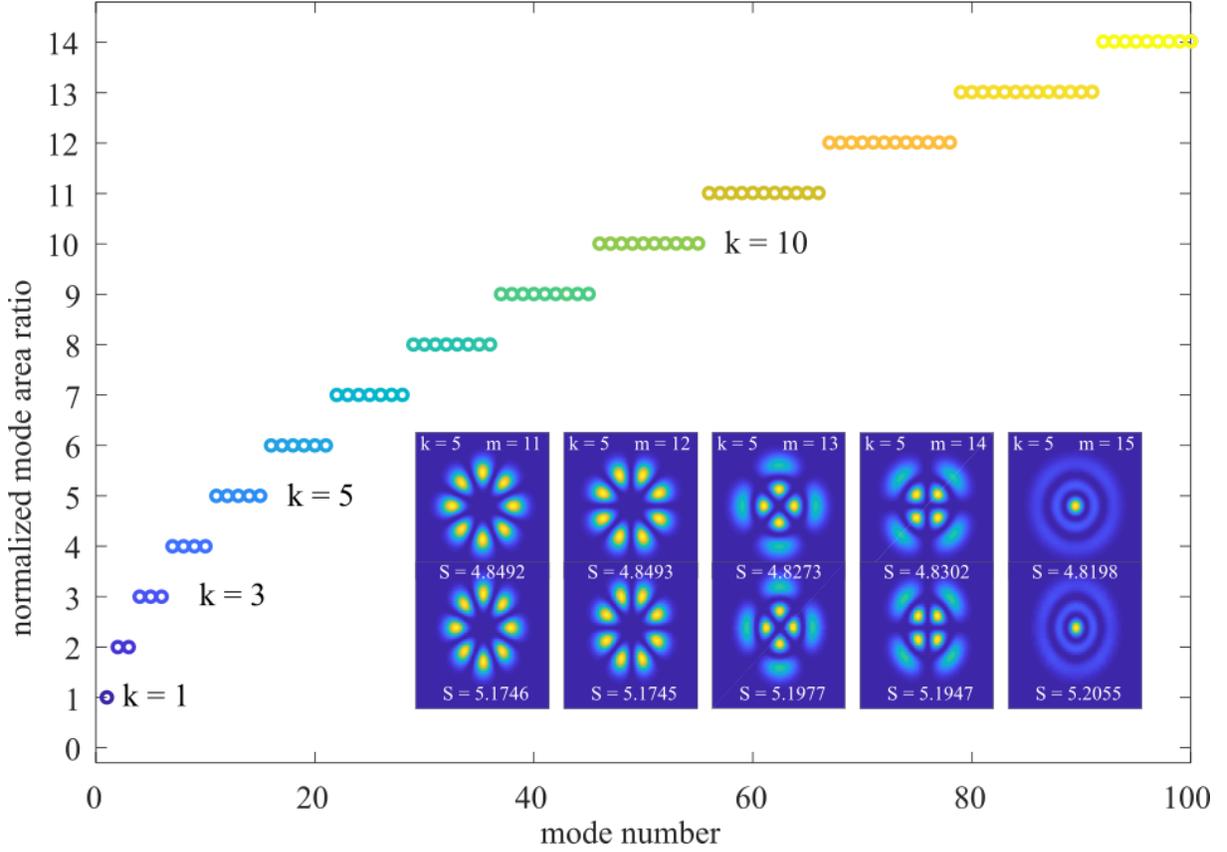

Figure S18. Normalized second-moment area ratio for the first 100 modes of the GRIN fiber. Inset shows the far-field (top row) and near-field (bottom row) intensity profiles for the modes in mode group 5 with corresponding normalized FF and NN second moment area.

**Section 17: Internal energy and average mode number**

For a short piece of GRIN fiber with $\Delta \ll 1$ and $\alpha \approx 2$, which is the case for our experiments, the propagation constant $\beta_g$ becomes a linear function of mode group number

$$\beta_k = n_0 k_0 - \frac{\sqrt{2\Delta}}{a} k. \tag{S16}$$

And the normalized propagation constant is given by

$$\varepsilon_k = -2k \tag{S17}$$

By assuming the spectral bandwidth is narrow and ignoring the wavelength dependence of the refractive index we can show that $\beta_k$ only depends on the mode group number (Fig. S19). As a result, we can approximate the internal energy with the average mode group number, which we have shown is conserved in our experiment. Thus, under these conditions we can write

$$\sum_{k=1}^{N} k g_k |c_k|^2 \propto \sum_{k=1}^{N} \varepsilon_k g_k |c_k|^2 = \sum_{i=1}^{M} \varepsilon_i |c_i|^2 \tag{S18}$$

where the first value shows the average mode group number, and second shows the internal energy. N is the total number of mode groups, and M is the total number of spatial modes. In the theory of optical thermodynamics, we are interested in the redistribution of energy among the modes with different propagation constants.

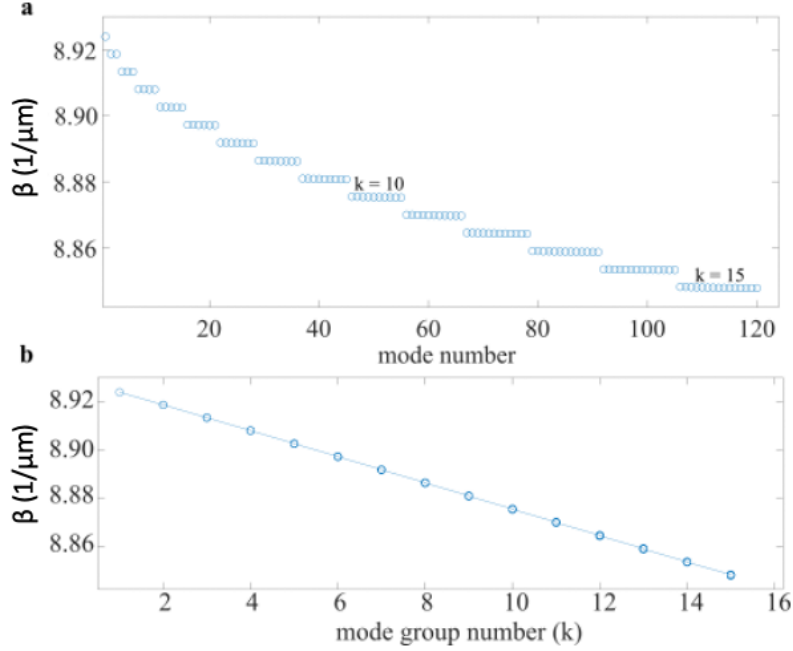

Figure S19. Propagation constant for first 120 modes of 50-μm GRIN fiber organized by mode number (**a**) and mode group number (**b**).

## Section 18: Beam quality and $M^2$

Consider a Gaussian beam with intensity pattern
$$I(x,z) = const \times \exp\left[-\frac{2x^2}{w^2(z)}\right] \tag{S19}$$
with $w(z) = w_0$ the minimum beam waist. The corresponding angular spatial-frequency power distribution is
$$I(k_x) = const \times \exp\left[-\frac{w_0^2 k_x^2}{2}\right] \tag{S20}$$
The variances of these gaussian-beam distributions, $I(x,z)$ and $I(k_x)$, are given by
$$\sigma_x(z) = \frac{w(z)}{2} \tag{S21a}$$
$$\sigma_{x0} = \frac{w_0}{2} \tag{S21b}$$
$$\sigma_{k_x} = \frac{1}{w_0} \tag{S21c}$$
where $\sigma_{x0}$ is the minimum spatial variance along $z$, which is located at the beam waist of the Gaussian beam. The product of the last two gives $\sigma_x \sigma_{k_x} = 1/2$.

Consider an arbitrary intensity distribution in the spatial $I(x,y,z)$ and frequency $I(k_x,k_y)$ domains. The variances $\sigma_x^2(z)$ and $\sigma_{k_x}^2$ are given by the following definitions:

$$\sigma_x^2(z) = \frac{\iint (x-\langle x \rangle)^2 I(x,y,z) dx dy}{\iint I(x,y,z) dx dy} = \langle x^2 \rangle(z) - \langle x \rangle^2(z) \tag{S22a}$$

$$\sigma_{k_x}^2 = \frac{\iint (k_x - \langle k_x \rangle)^2 I(k_x,k_y) dk_x dk_y}{\iint I(k_x,k_y) dk_x dk_y} = \langle k_x^2 \rangle - \langle k_x \rangle^2 \tag{S22b}$$

The variance $\sigma_x^2(z)$ always has a minimum value $\sigma_{x0}^2$ at some waist position. The beam quality factor along the x-axis $M_x^2$ is defined by:

$$M_x^2 = 2\sigma_{x0}\sigma_{k_x} \geq 1 \tag{S23}$$

Similarly, we can also define the beam quality factor along the y-axis by:

$$M_y^2 = 2\sigma_{y0}\sigma_{k_y} \geq 1 \tag{S24}$$

The cylindrical beam quality factor $M_r^2$ can also be defined:

$$M_r^2 = \sigma_{r0}\sigma_{k_r} \tag{S25}$$

where

$$\sigma_{r0}^2 = \min[\sigma_x^2(z) + \sigma_y^2(z)] \tag{S26a}$$

$$\sigma_{k_r}^2 = \sigma_{k_x}^2 + \sigma_{k_y}^2 \tag{S26b}$$

Note that a factor of 2 arises from normalization differences in 1-dimensional and 2-dimensional systems. When an optical beam is composed of more than one low-order or high-order Gaussian mode, the beam quality depends on the relative phases between the modes[5]. For example, an optical beam that consists of two Hermite-Gaussian modes of the order (0,0) and (2,0) can be written as:

$$\psi = c_{0,0}\Phi_{0,0} + c_{2,0}\Phi_{2,0} \tag{S27}$$

In this case, assuming the beam waists of the two modes are at the same position, the variance can be found according to Eq. (S22) and Eq. (S26) (some useful integrals are listed in the Appendix)

$$\begin{aligned}\sigma_{r0}^2 &= \langle \psi | x^2 + y^2 | \psi \rangle \\ &= \langle c_{0,0}\Phi_{0,0} + c_{2,0}\Phi_{2,0} | x^2 + y^2 | c_{0,0}\Phi_{0,0} + c_{2,0}\Phi_{2,0} \rangle \\ &= |c_{0,0}|^2 \langle \Phi_{0,0} | x^2 + y^2 | \Phi_{0,0} \rangle + |c_{2,0}|^2 \langle \Phi_{2,0} | x^2 + y^2 | \Phi_{2,0} \rangle + c_{0,0}c_{2,0}^* \langle \Phi_{0,0} | x^2 + y^2 | \Phi_{2,0} \rangle + c.c. \\ &= \frac{w_0^2}{2}\left(|c_{0,0}|^2 + 3|c_{2,0}|^2\right) + c_{0,0}c_{2,0}^* \langle \Phi_{0,0} | x^2 + y^2 | \Phi_{2,0} \rangle + c.c. \end{aligned} \tag{S28a}$$

$$\begin{aligned}\sigma_{k_r}^2 &= \langle \psi | k_x^2 + k_y^2 | \psi \rangle \\ &= \langle c_{0,0}\Phi_{0,0} + c_{2,0}\Phi_{2,0} | k_x^2 + k_y^2 | c_{0,0}\Phi_{0,0} + c_{2,0}\Phi_{2,0} \rangle \\ &= |c_{0,0}|^2 \langle \Phi_{0,0} | k_x^2 + k_y^2 | \Phi_{0,0} \rangle + |c_{2,0}|^2 \langle \Phi_{2,0} | k_x^2 + k_y^2 | \Phi_{2,0} \rangle + c_{0,0}c_{2,0}^* \langle \Phi_{0,0} | k_x^2 + k_y^2 | \Phi_{2,0} \rangle + c.c. \\ &= \frac{2}{w_0^2}\left(|c_{0,0}|^2 + 3|c_{2,0}|^2\right) + c_{0,0}c_{2,0}^* \langle \Phi_{0,0} | k_x^2 + k_y^2 | \Phi_{2,0} \rangle + c.c. \end{aligned} \tag{S28b}$$

The above equations show that the variances depend, not only on the amplitudes of the projection coefficients, but also on the relative phases between them. The cross-terms reach their extrema when the two modes are either in or out of phase, and becoming zero when the relative phase is $\pi/2$. A large-diameter gaussian beam with $M^2 = 1$, created by a superposition of several higher order modes of a parabolic fiber, is a particular example when the interference terms go to their overall extrema. However, if the two components are incoherently superposed, the cross terms vanish after ensemble averaging. In this case, the cylindrical beam quality $M_r^2$ can be found as:

$$M_r^2 = \sigma_{r0}\sigma_{k_r} = |c_{0,0}|^2 + 3|c_{2,0}|^2 \tag{S29}$$

More generally, when an optical field in a laser cavity or fiber system is an incoherent superposition of the eigenmodes (for example, Hermite-Gaussian modes $\Phi_{m,n}$, with projection coefficients $c_{m,n}$) the cylindrical quality factor $M_r^2$ is found to be[6]

$$M_r^2 = \frac{\sum_{m=0}^{\infty}\sum_{n=0}^{\infty}(m+n+1)|c_{m,n}|^2}{\sum_{m=0}^{\infty}\sum_{n=0}^{\infty}|c_{m,n}|^2} \quad (S30)$$

In this case, the beam quality factor is the same as the average mode number,

$$M_r^2 = \frac{\overline{U}}{\overline{U_1}} = \bar{k} \quad (S31)$$

The origin of the above relation is the fact that in each eigenmode of a harmonic oscillator, the kinetic energy is equal to the potential energy, so that $a + b = 2ab$, when $a = b$.

### Section 19: Linking $M^2$ and $U$ in the laboratory

From Eqs. (S13), (S15), (S22ab) and (S26ab), we know that $S_{NF}$ and $S_{FF}$ share similar definitions with $\sigma_r^2$ and $\sigma_{k_r}^2$

$$\frac{S_{NF}}{S_{NF}^1} = \frac{\iint_{-\infty}^{+\infty}(X^2+Y^2)|\bar{\psi}(X,Y)|^2 dXdY}{\iint_{-\infty}^{+\infty}(X^2+Y^2)|\overline{\Phi_1}(X,Y)|^2 dXdY} = \frac{\sigma_r^2}{\sigma_{r,1}^2} \quad (S32a)$$

$$\frac{S_{FF}}{S_{FF}^1} = \frac{\iint_{-\infty}^{+\infty}(K_x^2+K_y^2)|\bar{\psi}(K_x,K_y)|^2 dK_xdK_y}{\iint_{-\infty}^{+\infty}(K_x^2+K_y^2)|\overline{\Phi_1}(K_x,K_y)|^2 dK_xdK_y} = \frac{\sigma_{k_r}^2}{\sigma_{k_r,1}^2} \quad (S32b)$$

where $\sigma_{r,1}^2 = w_0^2/2$ and $\sigma_{k_r,1}^2 = 2/w_0^2$ is the variance of the fundamental mode in the local and frequency domains, respectively. According to Eq. (S25) the beam quality factor

$$M_r^2 = \sigma_r \sigma_{k_r} = \sqrt{\sigma_r^2 \sigma_{k_r}^2} = \sqrt{\frac{\sigma_r^2}{\frac{w_0^2}{2}} \times \frac{\sigma_{k_r}^2}{\frac{2}{w_0^2}}} = \sqrt{\frac{\sigma_r^2}{\sigma_{r,1}^2}\frac{\sigma_{k_r}^2}{\sigma_{k_r,1}^2}} = \sqrt{\frac{S_{NF}}{S_{NF}^1}\frac{S_{FF}}{S_{FF}^1}} \quad (S33)$$

In other words, by comparing with Eq. (16), we that if

$$\frac{S_{NF}}{S_{NF}^1} \sim \frac{S_{FF}}{S_{FF}^1} \quad (S34)$$

then $M_r^2 \sim \bar{n}$. This result indicates that, during thermalization process, the beam quality does not change, even though beam self-cleaning effects occur.

### Section 20: Measurements of the Hamiltonian

To estimate the linear Hamiltonian in the experiments, we need the propagation constants of the modes of the parabolic fiber. These are obtained by solving Eq. (S1) but without the nonlinear term,

$$i\frac{\partial E}{\partial Z} + \frac{1}{2k_0 n_0}\frac{\partial^2 E}{\partial X^2} + \frac{1}{2k}\frac{\partial^2 E}{\partial Y^2} - \frac{k_0 n_0 \Delta}{a^2}(X^2+Y^2)E = 0 \quad (S35)$$

The propagation constant of the $k^{th}$ mode group is given by[7]

$$\beta_k[\text{m}^{-1}] = -k\frac{\sqrt{2\Delta}}{a} \quad (S36)$$

For the fiber used in the experiments shown in Fig. 2, $a = 25$ μm and $\Delta \approx 0.01$. For the maximum input power level (52 kW), the modal occupancy distribution shown in Fig. 2 in the main text indicates that the internal energy (linear Hamiltonian) $U = -\sum_j |c_j|^2 \beta_j$ is about $1.5 \times 10^4$ kW/cm. The nonlinear Hamiltonian can be directly obtained by integrating the near-field intensity pattern captured by the camera:

$$H_{\mathrm{NL}} = \frac{k_0 n_2}{2} \iint dxdy\, [I(x,y)]^2 \tag{S37}$$

where $I(x,y)$ is the intensity distribution, $k_0$ is the wave vector in vacuum and $n_2 \approx 3.2 \times 10^{-20}$ W/m$^2$ is the Kerr nonlinear coefficient. In our experiments, the nonlinear Hamiltonian $H_{\mathrm{NL}}$ is approximately 3 kW/cm, i.e., it is at least three orders of magnitude smaller than the linear Hamiltonian. Therefore, the system operates in the weakly-nonlinear regime.

The optical spectrum measured over the range of peak powers used in the experiments is shown in Fig. S19a and b. The values of the Hamiltonian determined by inserting measured values in the expression $H = \sum \beta_i |c_i|^2$ and from the near- and far-field beam profiles are plotted in Fig. S19c.

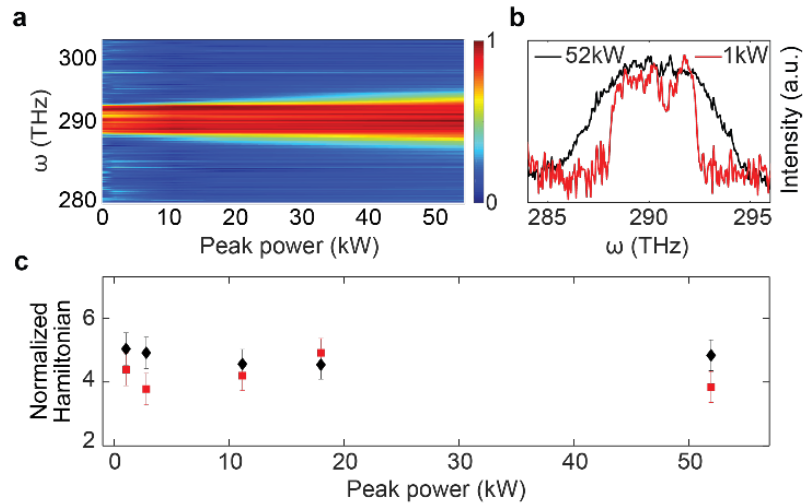

Figure S19. Conservation of internal (Hamiltonian) energy $H$. **a,** Measured power spectrum at the output of the 0.5-m multimode fiber *versus* input peak power. **b,** Power spectra recorded with minimum and maximum input powers indicate modest spectral broadening that results from nonlinear interactions that underlie thermalization of the spatial degrees of freedom. **c,** The normalized Hamiltonian energy $H/P$ remains invariant, as determined from the modal occupancies (black) and inferred from near- and far-field beam profiles (red). The invariance of the Hamiltonian $H$ during propagation is necessary for establishing the thermodynamic conditions needed to observe the RJ distribution. Error bars represent uncertainties as determined in the Supplementary Information.

**Section 21: Numerical simulations of thermalization in a step-index fiber**

Thermalization of light propagating in a step-index fiber is also possible. However, it typically requires an optical power at the ~1-MW level. To demonstrate these aspects, we perform simulations in a step-index fiber having a core radius of 25 μm and an NA of 0.206, under CW conditions. In the simulations, 42 out of 234 modes are randomly chosen and evenly excited. Figure S20 shows that after 10 m of propagation, when the peak optical power is increased to ~4 MW, the modal occupancies reach a perfect RJ distribution. For example, thermal equilibrium can also be obtained at 3 MW of peak power after propagating 20 m (Fig. S20f).

In practice, stimulated Raman scattering (which is neglected in this simulation with CW light) would be very strong at such high power and this would preclude experimental observation of thermalization. Self-focusing collapse would also be a concern at these power levels.

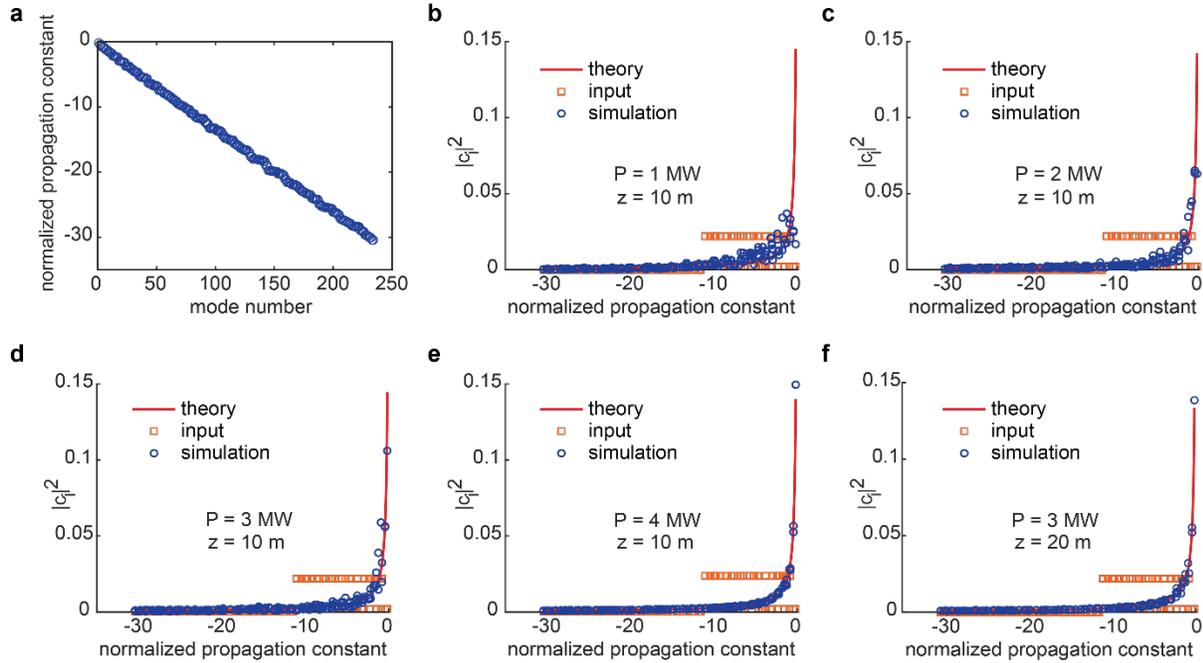

Figure S20. **a**. Normalized propagation constants of the fiber used in the simulations. **b-e**. Modal occupancies after 10 m of propagation under different power levels as indicated in the figure. **e-f**. light in a step-index fiber could reach thermal equilibrium after propagating 10 m @ 4 MW or 20 m @ 3 MW.

# Appendix

Physicists' Hermite polynomials are used throughout this Supplementary Information:
$$H_0(x) = 1,$$
$$H_1(x) = 2x,$$
$$H_2(x) = 4x^2 - 2,$$
$$H_3(x) = 8x^3 - 12x, \ldots \tag{A1}$$

Power normalized Hermite-Gaussian modes at beam waist (2D):
$$\Phi_{m,n}(x,y)e^{i\varepsilon_{mn}z} = \phi_m(x) \times \phi_n(y)e^{i\varepsilon_{mn}z}$$

$$= \left(\sqrt{\frac{1}{w_0}}\sqrt{\frac{2^{\frac{1}{2}-m}}{\pi^{\frac{1}{2}}m!}}H_m\left(\frac{\sqrt{2}x}{w_0}\right)e^{-\frac{x^2}{w_0^2}}\right) \times \left(\sqrt{\frac{1}{w_0}}\sqrt{\frac{2^{\frac{1}{2}-m}}{\pi^{\frac{1}{2}}m!}}H_n\left(\frac{\sqrt{2}y}{w_0}\right)e^{-\frac{y^2}{w_0^2}}\right)e^{i\varepsilon_{mn}z} \tag{A2}$$

Where $\varepsilon_{mn} = -2(m+n+1)$

Second moments of Hermite-Gaussian modes in space:
$$\sigma_x^2 = \langle \phi_m | x^2 | \phi_m \rangle = \frac{(1+2m)w_0^2}{4} \tag{A3}$$

Fourier transform of Hermite-Gaussian modes:
$$\phi_m(k_x) = \left(i^m \sqrt{\frac{w_0}{2}} \sqrt{\frac{2^{\frac{1}{2}-m}}{\pi^{\frac{1}{2}}m!}} H_m\left(\frac{kw_0}{\sqrt{2}}\right) e^{-\frac{k^2 w_0^2}{4}}\right) \tag{A4}$$

Second moments of Hermite-Gaussian modes in spectrum:
$$\sigma_{k_x}^2 = \langle \phi_m | k^2 | \phi_m \rangle = \frac{(1+2m)}{w_0^2} \tag{A5}$$